\documentclass[final,5p,times,twocolumn]{elsarticle}
\usepackage{multirow}
\usepackage{amssymb}
\usepackage{booktabs}
\usepackage{tabularx}
\usepackage{array}
\usepackage{longtable}
\usepackage{color}
\journal{Information Fusion}

\begin{document}

\begin{frontmatter}

\title{The Survey on Multi-Source Data Fusion in Cyber-Physical-Social Systems:\\ Foundational Infrastructure for Industrial Metaverses and Industries 5.0}

\author[1]{Xiao Wang\corref{cor1}}


\ead{xiao.wang@ahu.edu.cn}

\author[2]{Yutong Wang}

\ead{yutong.wang@ia.ac.cn}
\author[2,3]{Jing Yang}

\ead{yangjing2020@ia.ac.cn}

\author[4]{Xiaofeng Jia}
\ead{jiaxf@jxj.beijing.gov.cn}

\author[5]{Lijun Li}
\ead{lilijun@seu.edu.cn}

\author[6]{Weiping Ding}

\ead{dwp9988@163.com}

\author[2,3,7]{Fei-Yue Wang}
\ead{feiyue.wang@ia.ac.cn}

\cortext[cor1]{Corresponding author at: School of Artificial Intelligence, Anhui University, Hefei 266114, China}

\address[1]{School of Artificial Intelligence, Anhui University, Hefei 266114, China}
\address[2]{Institute of Automation, Chinese Academy of Sciences, Beijing 100190, China}
\address[3]{School of Artificial Intelligence, University of Chinese Academy of Sciences, Beijing 100049, China}
\address[4]{National Engineering Laboratory for Big Data Collaborative Security Technology, Beijing 100015, China}
\address[5]{ School of Automation, Southeast University, Nanjing, 210018, China}
\address[6]{School of Information Science and Technology, Nantong University,Nantong 226019,China}
\address[7]{ Faculty of Innovation Engineering, Macau University of Science and Technology, Macao 999078, China}





\begin{abstract}

As the concept of Industries 5.0 develops, industrial metaverses are expected to operate in parallel with the actual industrial processes to offer “Human-Centric” Safe, Secure, Sustainable, Sensitive, Service, and Smartness ``6S" manufacturing solutions. Industrial metaverses not only visualize the process of productivity in a dynamic and evolutional way, but also provide an immersive laboratory experimental environment for optimizing and remodeling the process. Besides, the customized user needs that are hidden in social media data can be discovered by social computing technologies, which introduces an input channel for building the whole social manufacturing process including industrial metaverses. This makes the fusion of multi-source data cross Cyber-Physical-Social Systems (CPSS) the foundational and key challenge. This work firstly proposes a multi-source-data-fusion-driven operational architecture for industrial metaverses on the basis of conducting a comprehensive literature review on the state-of-the-art multi-source data fusion methods. The advantages and disadvantages of each type of method are analyzed by considering the fusion mechanisms and application scenarios. Especially, we combine the strengths of deep learning and knowledge graphs in scalability and parallel computation to enable our proposed framework the ability of prescriptive optimization and evolution. This integration can address the shortcomings of deep learning in terms of explainability and fact fabrication, as well as overcoming the incompleteness and the challenges of construction and maintenance inherent in knowledge graphs. The effectiveness of the proposed architecture is validated through a parallel weaving case study. In the end, we discuss the challenges and future directions of multi-source data fusion cross CPSS for industrial metaverses and social manufacturing in Industries 5.0.

\end{abstract}

\begin{keyword}
Multi-source data fusion, CPSS, industrial metaverses, parallel manufacturing, social manufacturing

\end{keyword}

\end{frontmatter}
\newcommand\blfootnote[1]{%
  \begingroup
  \renewcommand\thefootnote{}\footnote{#1}%
  \addtocounter{footnote}{-1}%
  \endgroup
}

\section{Introduction}

While the popular conception of metaverses might involve social interactions, gaming, and entertainment in virtual spaces, industrial metaverses \cite{wang2023guest, chen2023metaverse, lee2022integrated} emphasizes improving business operations, enhancing productivity, streamlining manufacturing processes, and fostering innovation in industries. In essence, industrial metaverses can be described as a comprehensive digital counterpart that mirrors an entire manufacturing Cyber-Physical-Social System (CPSS) \cite{5552591,yilma2021systemic}, enabling interactions with its real-world counterpart and its surroundings, which allows those in decision-making roles to gain a clearer insight into historical events and predict future outcomes. Building upon and evolving from existing digital twin technologies \cite{tao2018digital, liu2021review}, industrial metaverses look beyond just efficiency and productivity. It emphasizes the significance and value of industry to society, fostering ``6S'' solutions in Industries 5.0 \cite{yao2022enhancing, wang2023steps}.

In general, industrial metaverses cover the entire value chain of manufacturing and move towards socialization. They bring together a CPSS encompassing stakeholders (from producers and consumers to designers), machinery (ranging from machinery equipment and robots to sensors), and components (like workpieces and raw materials). Collaborative design and the creation of a value chain that offers a competitive edge become the focus of attention, which could only be realized by anticipating the evolving diversity and unique requirements of future human societies\cite{ wang2018societies} by global innovation. Thus, a model that is capable of perception, prediction, and planning becomes an inevitable requirement, serving as the technical foundation for the operation of industrial metaverses.

However, manufacturing CPSS results in an overwhelming volume of multi-source data, which makes it difficult to extract valuable knowledge and make correct judgments on intrinsic complexities and unforeseen external events. Multi-source data in CPSS refers to data coming from different sources, including sensors in physical space, databases in cyber space, and social media in social space. Traditional data-mining techniques fall short in discerning and comprehending patterns in such data, necessitating the use of CPSS multi-source data fusion approaches \cite{wang2019data, meng2020survey, zhang2021tensor, yang2023representation}. Multi-source data fusion could efficiently integrate data from multiple sources to produce more consistent, accurate, and useful information than that provided by any single source alone \cite{zhang2021multi, khaleghi2013multisensor, chai2023heterogeneous}. Therefore, it can improve the quality of data, enabling more robust analyses, and supporting better decision-making in various domains such as clinical diagnosis \cite{ qu2023qnmf, nie2020multi, tao2022multi},  anomalies/outlier detection \cite{zhang2023multi,cai2014multi, li2018physics} and personalized recommendation \cite{guo2018mobile,cheng2019friend}.


\begin{figure}[ht]
\centering
\includegraphics[width=\linewidth]{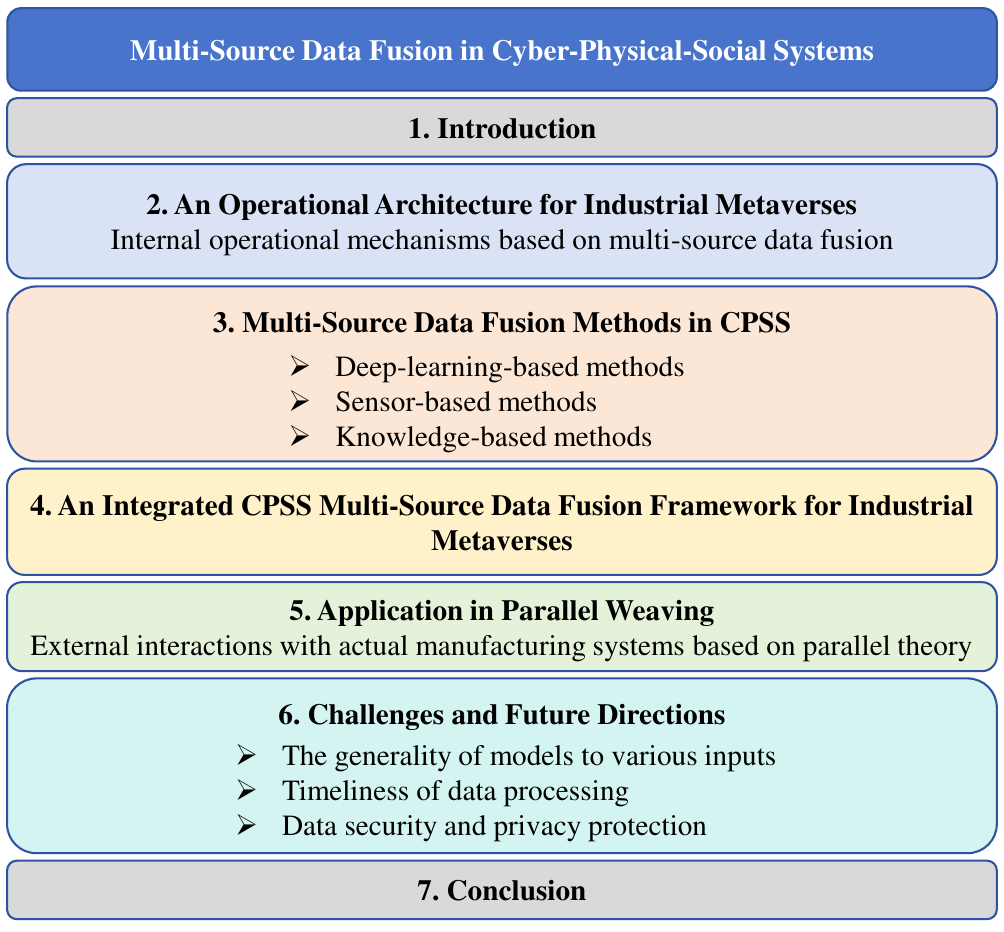}
\caption{The organizational layout of this paper.}
\label{Overall}
\end{figure}

\begin{figure*}[ht]
\centering
\includegraphics[width=0.85\linewidth]{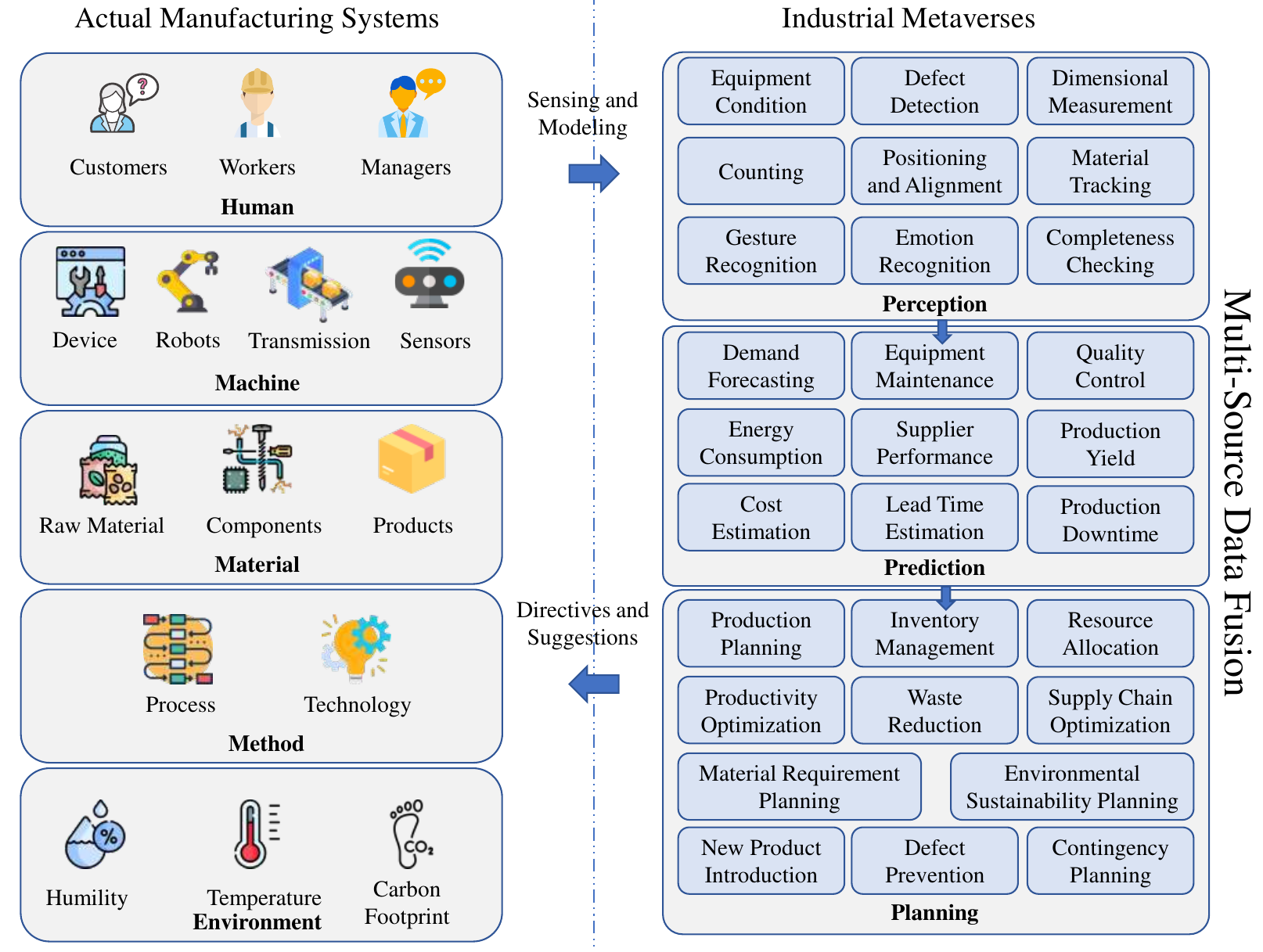}
\caption{A operational architecture for industrial metaverses.}
\label{Architecture}
\end{figure*}

Within the context of CPSS smart manufacturing under Industries 5.0, our research aims to provide a multi-source-data-fusion-based comprehensive solution for industrial metaverses. Though there are already plenty of works summarizing and reviewing multi-source data fusion methods, they mainly focus on a specific category of methods, such as \cite{wang2019data, zhang2021tensor} focus on tensor-based methods, \cite{smirnov2019knowledge, zhao2020multi} focus on knowledge-based methods, and \cite{10123038, han2023survey, guo2019deep} focus on deep learning methods. Lacking a comparison between different types of methods, these papers also fail to provide a solution for multi-source data fusion to tackle multi-task situations. Therefore, the main contributions of this paper are summarized as follows:
\begin{itemize}
    \item We introduce a ``6S" manufacturing solutions under Industries 5.0, which is a hierarchical operational architecture of industrial metaverses driven by a synergized CPSS multi-source data fusion framework. This framework combines foundation models (FMs) with knowledge graphs (KGs), and achieves a complementary integration of their strengths.
    
    \item We categorize the mainstream CPSS multi-source data fusion methods into three types based on data representation format, and provide a comprehensive comparison of these three types.
    
    \item We present the application of our proposed architecture on parallel weaving, executing collaborative design, customization, and distant interactions within industrial metaverses. These studies validate the feasibility and efficiency of the operational architecture we propose.
\end{itemize}

The organizational layout of the paper is illustrated in Fig. \ref{Overall} and its rest is organized as follows. Section 2 introduces a hierarchical operational architecture of industrial metaverses, describing its internal operational mechanisms based on multi-source data fusion methods, which serve as the technical foundation. Section 3 reviews the mainstream CPSS multi-source data fusion methods, divides them into three categories: deep-learning-based methods, tensor-based methods, and knowledge-based methods, and identifies the application scope, pros and cons of each category of methods. In Section 4, to compensate for the shortcomings of deep-learning-based and knowledge-based approaches, we propose a synergized CPSS multi-source data fusion framework, which improves perception, prediction, and planning performance. In Section 5, we apply our proposed architecture to parallel weaving, demonstrate interactions between industrial metaverses and actual manufacturing systems, and verify the effectiveness of our proposed architecture. In Section 6, we point out the current challenges and future directions. Finally, we conclude this paper in Section 7.

\section{An operational architecture for industrial metaverses}

Industry 4.0 focuses on automation, data exchange, and manufacturing technologies. Industry 4.0 is marked by the rise of smart factories, Internet of Things (IoT), Artificial Intelligence (AI), machine learning, and more. In contrast, Industries 5.0 shifts focus back to the human element in manufacturing and technology \cite{wang2023steps, wang2024human,zhang2023towards}. It emphasizes the collaboration between humans and smart systems, highlighting personalized production, sustainability, and the integration of human creativity and craftsmanship with the efficiency and precision of machines. Industries 5.0 seeks to balance automation with human touch, intuition, and creativity, aiming for a more sustainable and human-centric approach.

``Metaverse'' is a collective virtual shared space, created by the convergence of virtually enhanced physical reality and physically persistent virtual spaces \cite{wang2022metaverse, gadekallu2023blockchain, wang2022dao}. Serving as a powerful tool for realizing the goals of Industries 5.0 \cite{wang2022framework,wang2023new, wang2023chat}, Industrial metaverse involves using virtual and augmented reality, digital twins, and other immersive technologies to create a virtual representation of industrial processes, environments, and products. This allows for simulation and optimization, customization and personalization, remote monitoring and maintenance, and human-machine interaction, thereby enhancing human-centric manufacturing, fostering innovation, improving safety and training, promoting sustainability, and boosting overall efficiency and productivity.

When constructing industrial metaverses, all manufacturing resources that participate in or influence manufacturing activities are modeled \cite{qi2019smart, wang2021smart, garcia2022towards}, as shown in Figure \ref{Architecture}. Specifically, within a plant, the primary resources can be summarized as ``Human-Machine-Material-Method-Environment". These resources interact with each other to carry out a wide range of manufacturing activities. ``Human" refers to customers, workers and management personnel. ``Machine" encompasses equipment (like machine tools, robots, conveyor systems, etc.) for production and sensing devices (like various smart sensors, radio frequency identification, etc.) for perception. ``Material" includes raw materials, components, and products. ``Method'' denotes the procedures, techniques, and processes used in production. Additionally, ``Environment'' refers to both the physical environment (like temperature, humidity, and lighting conditions) and the broader organizational culture and working environment.

Within industrial metaverses, the operating systems consistently gather data on the status of the ``Human-Machine-Material-Method-Environment" components. Through data presentation, fusion, and analysis, models gain production knowledge and perform perception, prediction, and planning. Perception refers to the ability to gather data and information and have an understanding of the manufacturing environment in real time. This involves sensors, cameras, and other data-capturing devices that monitor various aspects of the manufacturing process. Perception systems collect data on things like equipment status, product quality, material flow, and environmental conditions. This data is essential for maintaining quality control, ensuring safety, and optimizing manufacturing processes. Prediction involves analyzing the data collected from perception systems. Predictive models can anticipate potential issues or deviations from normal operations in the manufacturing process. For example, they can predict equipment failures, quality defects, or supply chain disruptions. By forecasting these events, manufacturers can take proactive measures to avoid costly downtime, improve product quality, and optimize production schedules. Planning takes the insights gained from perception and prediction and uses them to optimize production plans and schedules. It involves making decisions about resource allocation, production sequencing, inventory management, and other factors to ensure that manufacturing operations run smoothly and efficiently. Planning systems can dynamically adjust production schedules based on real-time data and predictions to meet changing demands, minimize lead times, and optimize resource utilization.

\begin{table*}[ht]\footnotesize
\centering
\caption{{\normalsize A summary of multi-source data fusion methods in CPSS}}
\begin{tabularx}{1.0\linewidth}{p{2cm}p{2.5cm}p{3cm}p{4.5cm}p{4.5cm}}
     \label{tab:all_tool_urls} \\
    \toprule

    Data Fusion Methods & Input Data & Processing Steps & Pros & Cons\\
    \midrule
    Deep-learning-based methods & Unstructured data & Feature encoding, feature fusion, and feature decoding & Highly scalable; General knowledge & Only processing unstructured data; Data hungry; Implicit knowledge; Hallucination\\\hline
    Tensor-based methods & Unstructured data, semi-structured data, and structured data & Tensor representation, tensor fusion, and dimensionality reduction & Simple high-order data representation and fusion operation; Noise reduction & Limited in handling complex patterns\\\hline
    Knowledge-based methods & Unstructured data, semi-structured data, and structured data & Knowledge extraction, knowledge representation, knowledge reasoning, and knowledge fusion & Low data dimensionality; Interpretability; High accuracy; Highly scalable & Sophisticated knowledge extraction, representation, reasoning, and fusion process; Incompleteness; Domain expertise requirements\\
    \bottomrule
\end{tabularx}
\label{Comp}
\end{table*}

In summary, perception, prediction, and planning are integral components to improve the efficiency, quality, and flexibility of production processes, which play a crucial role for industrial metaverses. However, with the popularization of sensing technology \cite{andronie2021sustainable,jiang2020review,majid2022applications} and the development of communication technology \cite{aceto2019survey, colombo202170}, there is a vast amount of multi-source heterogeneous data generated in industrial metaverses, which poses a great challenge for data analytics to achieve perception, prediction, and planning.

\section{Multi-source data fusion methods in CPSS}

In this section, we provide a comprehensive overview of multi-source data fusion methods within CPSS \cite{yang2023parallel,yang2023parallelCPS, 9896794}, divide methods into three categories: deep-learning-based methods, tensor-based methods, and knowledge-based methods, and identify the application scope, pros and cons of each category of methods. A summary of multi-source data fusion methods in CPSS is shown in Table \ref{tab:all_tool_urls}. Deep learning methods rely on analyzing and processing large amounts of data to identify patterns, trends, and correlations without pre-existing knowledge or assumptions. Employing parallel computing, they work in an end-to-end architecture to extract insights and make predictions. The disadvantages include limited input data type and interpretability. Tensor-based methods exploit simple and intuitive tensor representation and operation to perform fusion. However, high data volume makes it necessary to perform dimensionality reduction, which leads to high computational complexity and memory requirements. Knowledge-based methods leverage existing knowledge, expertise, and domain-specific rules to construct KGs, and are often more interpretable. However, the knowledge extraction, representation, reasoning and fusion process could be sophisticated. And the incompleteness always leads to reasoning failures.

\subsection{Deep-learning-based multi-source data fusion methods}
In this section, we present an overview of multi-modal deep-learning-based data fusion methods in CPSS.  Multi-modal data fusion \cite{wang2021survey, wu2023mfir, huang2023multi} is thought of as a significant subdomain of multi-source data fusion where data is collected from diverse sources, but it all belongs to unstructured data. Specifically, multi-modal data fusion refers to integrating multiple-modality data such as text, images, audio, and video with a wealth of social information \cite{tian2023acf, wang2015social}. Though traditional machine learning methods such as Support Vector Machine (SVM) \cite{adams2003semantic} and Hidden Markov Model (HMM) \cite{bourlard1996mew} have long been used for multi-source data fusion \cite{dogan2021machine}, with the advent and rise of deep learning, we mainly focus on deep-learning-based multi-modal data fusion \cite{QU2023102172} for high performance.

Deep learning has emerged as a powerful tool for multi-modal data fusion, whose performance even surpass humans \cite{summaira2021recent}. Deep-learning-based methods for multi-modal data fusion leverage the powerful representation learning capabilities of deep neural networks, enabling the integration of information from multiple sources to create a unified representation. Furthermore, for multimodal data fusion methods in CPSS, we primarily focus on the vision-language domain, as both the data and tasks in this field are closely related to society. And the input data includes images, video, and text, while the downstream tasks include image-text tasks, video-text tasks and computer vision tasks.
In 2014-2015, researchers started exploring the idea of combining images and questions in a unified framework, and vision-language tasks such as visual question answering \cite{antol2015vqa}, and text grounding \cite{plummer2015flickr30k} began to take shape. Since then, the release of larger and more diverse datasets has significantly boosted the development of this field. These methods typically consist of a text encoder, an image feature extractor, a multimodal fusion module (often with attention), and a specific-task decoder. In this process, models generally use a Transformer architecture for encoder-decoder.

Motivated by the impressive performance of Transformers, researchers come to realize that without relying heavily on modality-specific architectures, Transformer-based multi-modal data fusion could perform multi-task learning such as visual question answering, image captioning, segmentation, and classification \cite{cai2023multi}. Thereby Transformer has become a mainstream approach for multi-modal data fusion. To address the issues of the inability to train recurrent network models in parallel, as well as the requirement for significant storage resources to memorize the entire sequence of information, Transformer was proposed for Natural Language Processing (NLP) in 2017 \cite{vaswani2017attention}, and have made significant strides on a variety of NLP tasks \cite{devlin2018bert, brown2020language}. Due to their proficiency in capturing global context and long-range dependencies, Transformers have started being employed in visual domains  \cite{dosovitskiy2020image, carion2020end} since 2020.

\begin{figure*}[ht]
\centering
\includegraphics[width=0.7\linewidth]{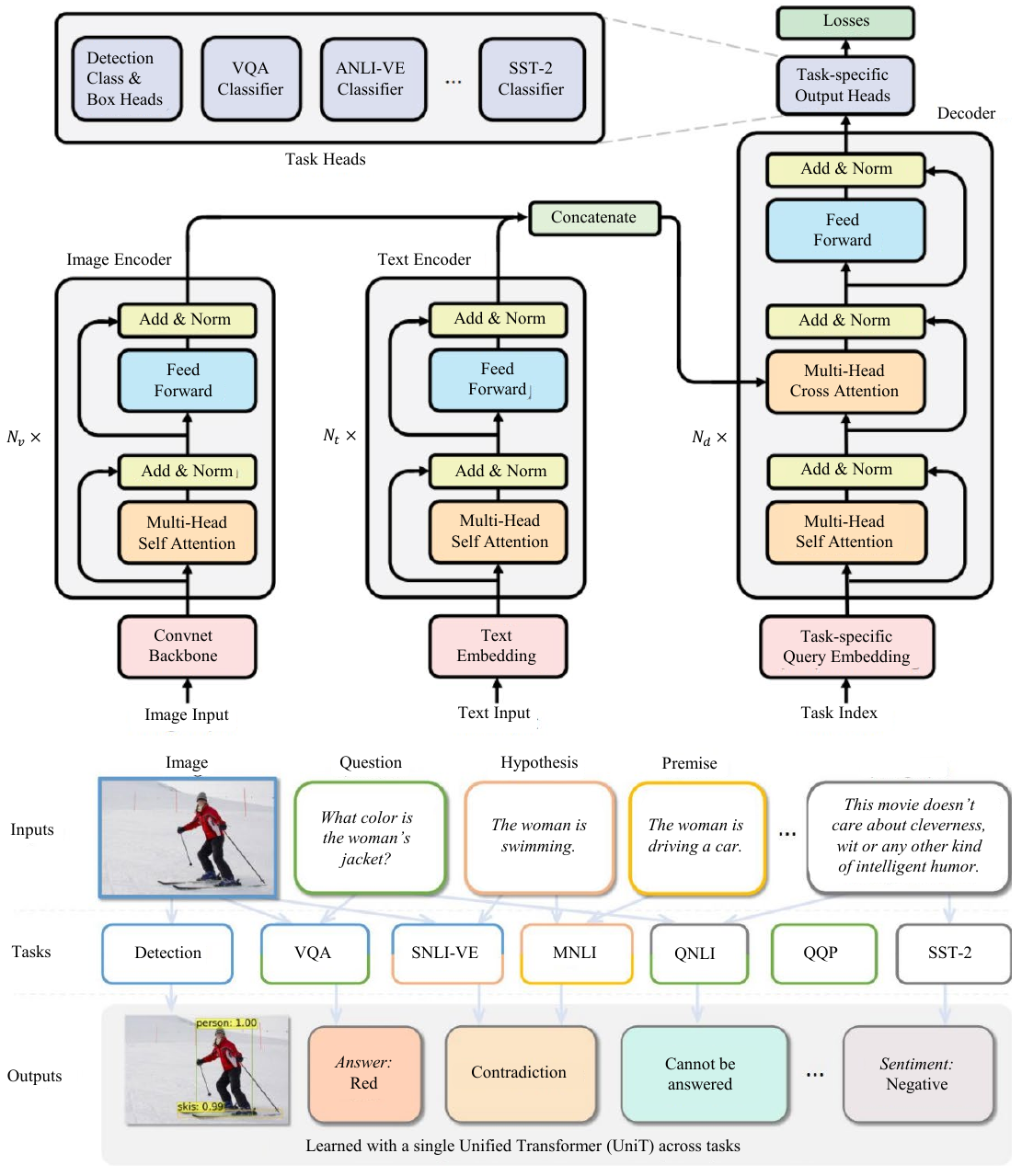}
\caption{An overview of UniT. UniT employs a unified transformer encoder-decoder architecture. It efficiently addresses a diverse range of tasks, encompassing object detection, vision-and-language reasoning, and natural language understanding \cite{hu2021unit}.}
\label{UniT}
\end{figure*}

Transformer has an encoder-decoder architecture. The encoder processes the input sequence, typically a sequence of embeddings in which positional encodings are added to provide positional information. It consists of multiple layers, each of which includes two sub-layers: a multi-head self-attention mechanism and a position-wise feed-forward neural network. The self-attention mechanism attends to all positions in the input sequence and computes a weighted sum of the embeddings, allowing the model to consider the context of each word. It computes a weighted sum of values based on the similarity between a query and a set of key-value pairs:
\begin{equation}
Attention(Q,K,V)=\mathrm{softmax}(\frac{QK^T}{\sqrt{d_k}})V
\end{equation}
where $Q$, $K$, and $V$ are obtained through linear transformations of the input embeddings, $d_k$ is the dimension of $Q$ and $K$. In the Transformer, multi-head attention is used, where multiple attention heads operate in parallel, capturing different types of dependencies. The position-wise feed-forward neural network comprises two linear transformations separated by a ReLU activation function:
\begin{equation}
FFN(x)=\max(0, xW_1+b_1)W_2+b_2.
\end{equation}

The decoder takes the output of the encoder and generates the output sequence. Similar to the encoder, it also consists of multiple layers, but with an additional sub-layer that performs multi-head attention over the encoder's output. In the visual domain, Convolutional Neural Network (CNN) features or patches could be used as tokens for Transformer input. For data fusion \cite{10123038, han2023survey, nguyen2023multimodal}, summation, concatenation, hierarchical attention, and cross-attention could be applied to different modalities.

Most Transformer-based multi-modal learning methods work on a single domain \cite{raffel2020exploring, lu202012, yoon2023multimedia}. Compared to single-domain tasks, multi-domain tasks are more in line with practical needs and should be given greater importance, where multi-domain tasks refer to tasks from three domains (visual recognition, language understanding, and multi-modal reasoning) \cite{dou2022coarse}. Hu and Singh \cite{hu2021unit} introduce UniT, a Unified Transformer model aiming at simultaneously mastering the most prominent tasks spanning various domains, as shown in Fig. \ref{UniT}. These tasks encompass a wide range from object detection to natural language comprehension and multi-modal reasoning. The UniT model employs one encoder to encode each input modality and utilizes a shared decoder to generate predictions for each task based on the concatenated encoded input representations. Task-specific output heads follow suit. The entire model undergoes end-to-end joint training, incorporating losses from each task. And in experimental evaluations, UniT is trained on eight datasets encompassing seven tasks, achieving exceptional performance on each task while utilizing substantially fewer parameters. Afterward, Singh et al. propose FLAVA \cite{singh2022flava}, which exploits a multi-modal encoder Transformer that combines the encoded unimodal image and text for multi-modal reasoning. Wang et al. propose ONE-PEACE \cite{wang2023one} to seamlessly align and integrate representations across vision, audio, and language modalities. 

Transformers are designed to be highly scalable, handling large amounts of data and a large number of parameters efficiently. With this robust and flexible architecture, Transformer becomes the backbone of the mainstream FMs (also known as large models) \cite{alayrac2022flamingo, awadalla2023openflamingo, gao2023llama}. Pre-trained on a vast and diverse dataset, FMs are designed to be adaptable and can be fine-tuned for a wide range of tasks across different domains with billions of parameters \cite{bommasani2021opportunities, li2023multimodal, xu2023multimodal, wang2023large}.

In summary, compared to traditional machine learning methods, deep-learning-based multimodal data fusion methods have the advantages of an end-to-end architecture, faster inference speed, and extremely high accuracy. Due to their powerful representation capability, deep learning methods can handle large-scale datasets and diverse modalities, and they can learn complex patterns and relationships between different modalities. Additionally, these methods currently work on a pretrain-finetune pipeline, which could make full use of unlabeled big data. Deep-learning-based multi-source data fusion methods hold great promise for advancing the state-of-the-art performance in multi-modal data fusion.

\subsection{Tensor-based multi-dource fata fusion}

Multi-modal data is from diverse sources, yet it falls under unstructured data. Multi-source heterogeneous data comprises structured, semi-structured, and unstructured data. Hence, we investigate tensor fusion that could fuse structured, semi-structured, and unstructured data together. 

Tensors are multi-dimensional arrays, and can be thought of as an extension of matrices to higher orders. As a result, tensors offer a natural and innate means of representing higher-order data, making them an ideal choice for solving the representation problem of multi-source heterogeneous data. Currently, tensors have been widely employed in the data fusion field for adaptive clustering, multi-modal prediction, and recommendation in CPSS. There are mainly two methods for tensor fusion: one is the unified tensor method, and the other is the transition tensor method. In this subsection, we mainly discuss these two methods on tensor representation, tensor fusion, and tensor analysis procedures.
\begin{figure}[ht]
	\centerline{\includegraphics[width=20.pc]{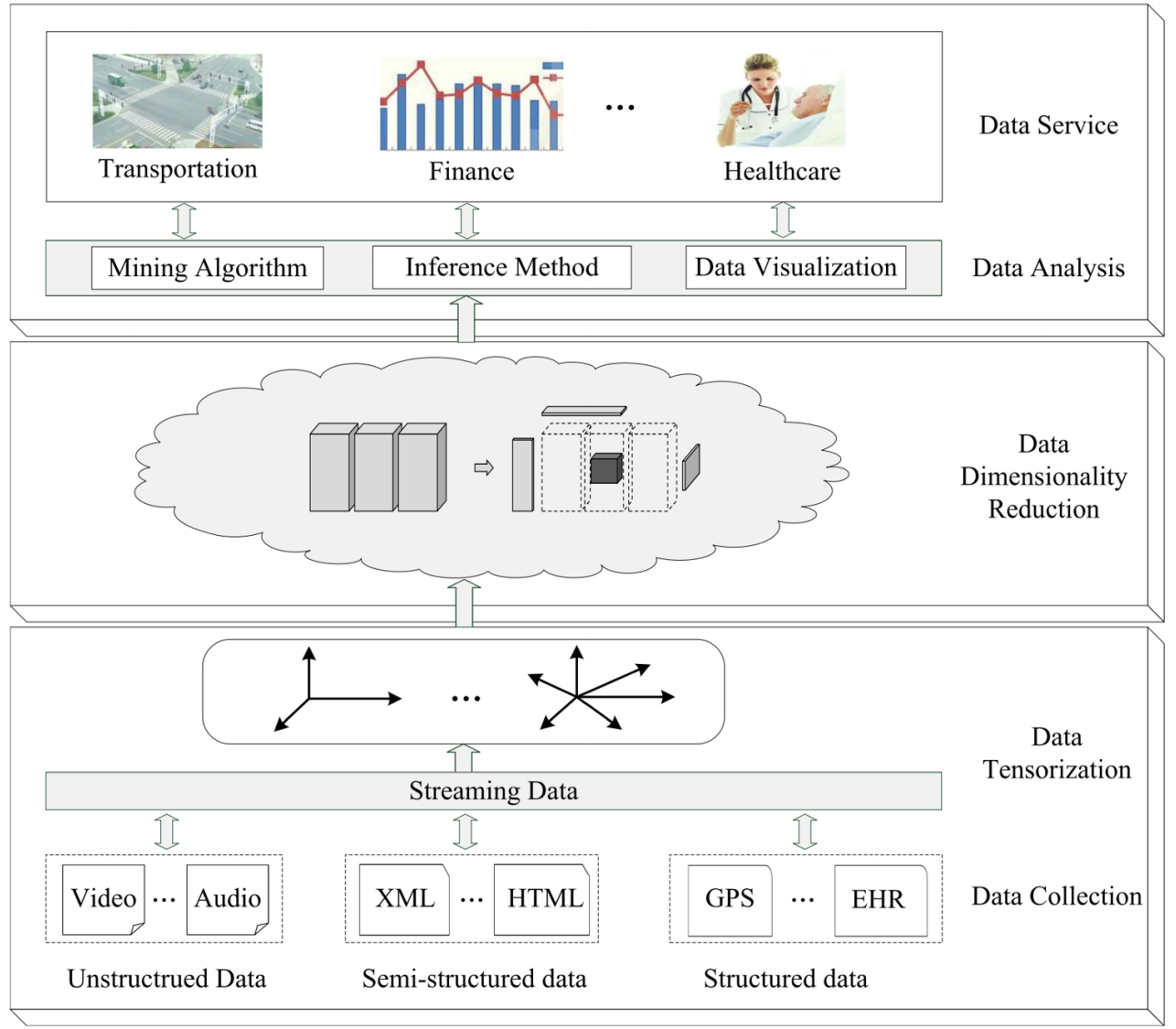}}
	\caption{A typical framework of unified tensor method \cite{kuang2014tensor}.}
	\label{tensorflow}
\end{figure}

\subsubsection{Unified tensor methods}

A typical framework of the unified tensor method is shown in Fig. \ref{tensorflow}. Due to the lack of uniformity among the collected data from CPSS, it is necessary to accurately represent unstructured, semi-structured, and structured data according to their original formats and integrate them together. These heterogeneous data is first transformed into low-order sub-tensors according to their original formats. Then tensor fusion is performed to integrate all the sub-tensors as a unified high-order tensor. When collecting data from various sources, two fundamental characteristics are time and space, and the primary recipients of data services are users. For this reason, a universal tensor-based data model has been established in \cite{kuang2014tensor}:

\begin{equation}
T \in \mathcal{R}^{I_t \times I_s \times I_u \times I_1 \times \cdots \times I_P}
\end{equation}
where $\mathcal{R}$ is defined in the real number domain, $\times$ is the Cartesian product. $T$ contains two parts, a base part and an extensible part. In the base part, $I_t$, $I_s$, and $I_u$ denote time, space, and user, respectively. In the extensible part, $I_1$, $\cdots$, $I_P$ denote the different data characteristics. Subsequent works such as \cite{kuang2016tensor, 7134729} still use this tensorization method. But \cite{7364232} augments the three-order base part into six-order tensor:
\begin{equation}
T_{base} \in \mathcal{R}^{I_{tim} \times I_x \times I_y \times I_z \times I_{dev} \times I_{usr}}
\end{equation}
where $I_{tim}$, $I_x$, $I_y$, $I_z$, $I_{dev}$, and $I_{usr}$ refer to time, latitude, longitude, altitude, cyber resource and user respectively.

In this way, data from physical space include sensor signals such as Global Positioning System (GPS), video, and audio. GPS data can be represented as a four-order tensor $\mathcal{R}^{I_{tim} \times I_x \times I_y \times I_z}$. Video clip with MPEG-4 format can be represented as a four-order tensor $\mathcal{R}^{I_{f} \times I_w \times I_h \times I_{cs}}$, where $I_f, I_w, I_h, I_{cs}$ refer to frame, width, height, and color space. An audio clip of WAV format can be expressed as a two-order tensor $\mathcal{R}^{I_l \times I_r}$ where $I_l$ and $I_r$ indicate the left and right soundtrack. Data from cyber space involves RDF and OWL data written in XML. And they can be tensorized as $\mathcal{R}^{I_{rt} \times I_{ci} \times I_{ci} \times I_{ne}}$, where $I_{rt}$ and $I_{ci}$, $I_{ne}$ denote relationship, class or individual, and character encoding. Data from social space like social relationships between users can be formalized as a three-order tensor $\mathcal{R}^{I_{usr} \times I_{usr} \times I_{rel}}$, where $I_{usr}$, $I_{rel}$ indicate user and social relationship.

\begin{figure}[ht]
	\centerline{\includegraphics[width=18.pc]{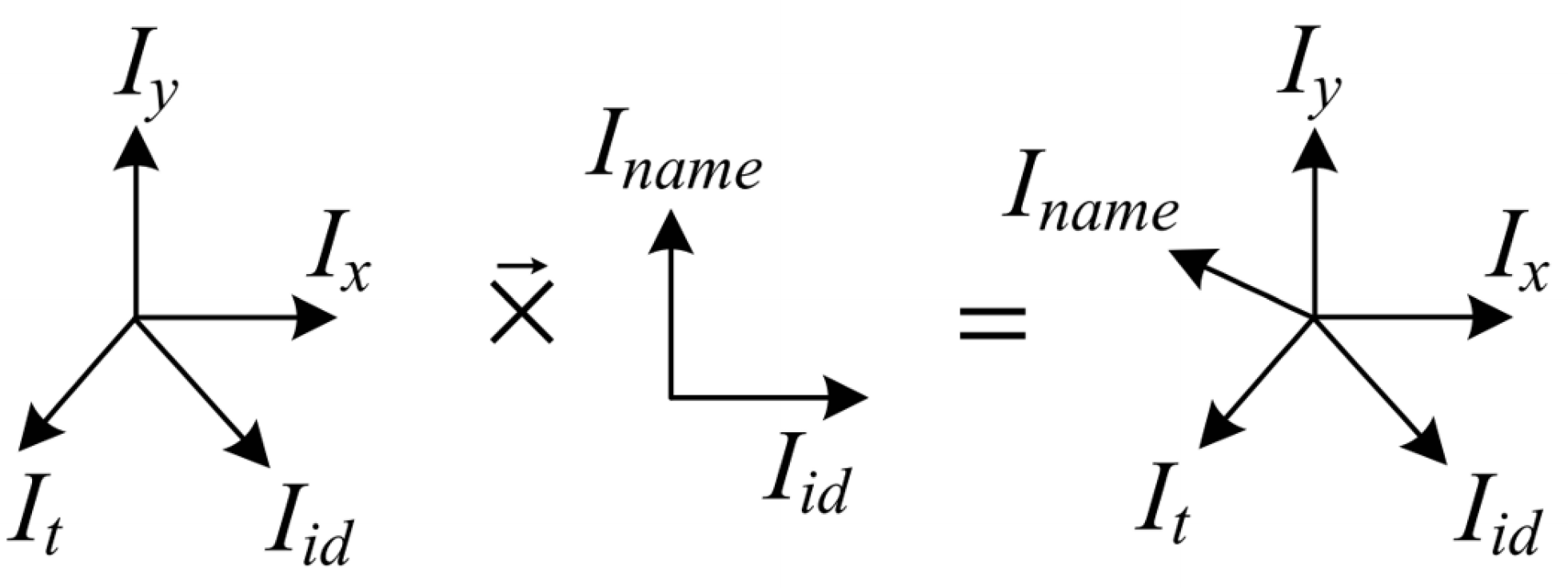}}
	\caption{Two sub-tensors are unified as a five-order tensor by tensor extension operator \cite{kuang2014tensor}.}
	\label{tensor-fusion}
\end{figure}

For tensor fusion, the tensor extension operator, as depicted in Fig. \ref{tensor-fusion}:
\begin{equation}
f: A \overrightarrow{\times}  B \rightarrow C, C \in \mathcal{R}^{I_t \times I_s \times I_u \times I_1 \times I_2}
\end{equation}
where $A \in \mathcal{R}^{I_t \times I_s \times I_u \times I_1}$, $B \in \mathcal{R}^{I_t \times I_s \times I_u \times I_2}$. $\overrightarrow{\times}$ operator merges the identical orders with elements accumulated together, while keeping the diverse orders.

With the tensor fusion operation, a unified tensor has too many orders, and may increase the time and space complexity for following analysis. In addition, the unified tensor generated from heterogeneous tensor are usually sparse. So dimensionality reduction \cite{abed2022contemporary} is applied to process the high-dimension tensorized data, and extract the core data with less orders for communication, computation, and storage. High-order singular value decomposition (HOSVD) \cite{kolda2009tensor} is first proposed to perform tensor unfolding, core tensor construction, and approximation tensor construction. And the reconstructed data could be more accurate than the original data as the inconsistency, noise and redundancy are removed. Inspired by HOSVD, a dimensionality reduction method called Incremental High-Order Singular Value Decomposition (IHOSVD) is introduced in \cite{kuang2014tensor}. This method involves the iterative application of the incremental matrix decomposition algorithm to update the truncated unitary orthogonal bases and derive the updated core tensor. The result is a condensed representation of the original high-order sparse unified tensor. For high decomposition efficiency, Lanczos-based high-order orthogonal tensor singular value decomposition algorithms are proposed in \cite{kuang2014tensor, wang2019ho}. During each iteration, the Lanczos method is used for solving eigenvalues and eigenvectors of the large-scale sparse matrix. Additionally, some works \cite{feng2018privacy, feng2019practical} combine the properties of homomorphic encryption and tensor computation for constructing privacy-preserving Lanczos schemes in the integrated edge-fog-cloud-based smart environment. 

\begin{figure}[ht]
	\centerline{\includegraphics[width=20.pc]{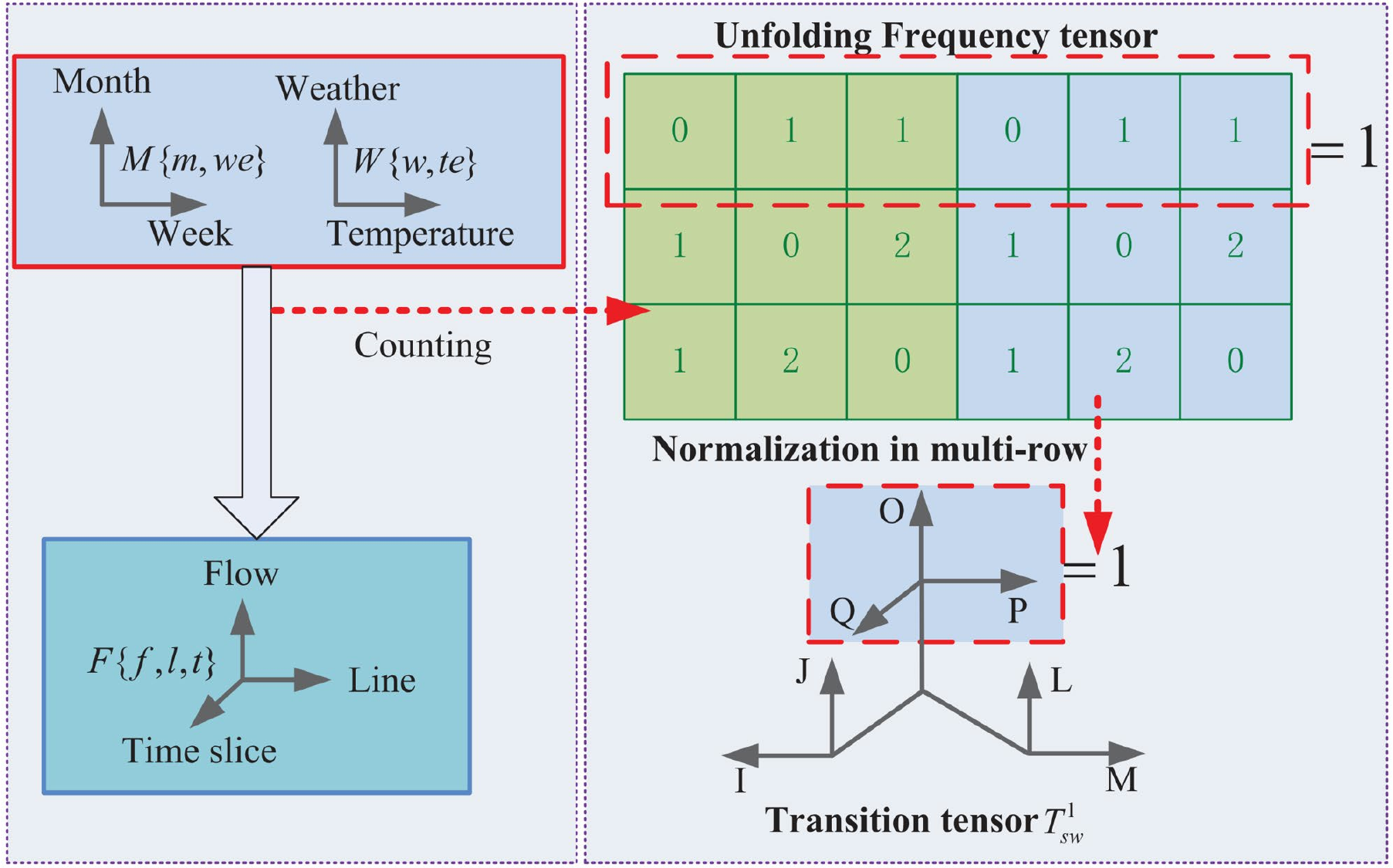}}
	\caption{The process of generating the multi-space transition tensor \cite{wang2018cyber}.}
	\label{Transition}
\end{figure}

\subsubsection{Transition tensor methods}

Unlike the unified tensor method that directly represents data, the transition tensor method represents the state transition probability. A multivariate transition tensor is first constructed that fuses time, location, and social information data \cite{8701474}. To obtain a spatio-temporal transition tensor from GPS data, the point of interest (POI) sequence is first extracted, and the temporal information is divided into discrete time intervals. A state is, therefore,
introduced as $S=\{\{T_1,P_1\}, \cdots, \{T_I,P_J\}\}$, where $T_i$ refers to the $i$th temporal state, $P_j$ refers to the $j$th spatio state, and $I$ and $J$ are the total number of states for time, and space respectively. With this state space, it is possible to construct a Markov chain and generate the corresponding transition tensor:
\begin{equation}
T=\frac{count((T_1,P_1)\rightarrow(T_2,P_2))}{count((T_1,P_1)}.
\end{equation}

\begin{figure}[ht]
	\centerline{\includegraphics[width=18.pc]{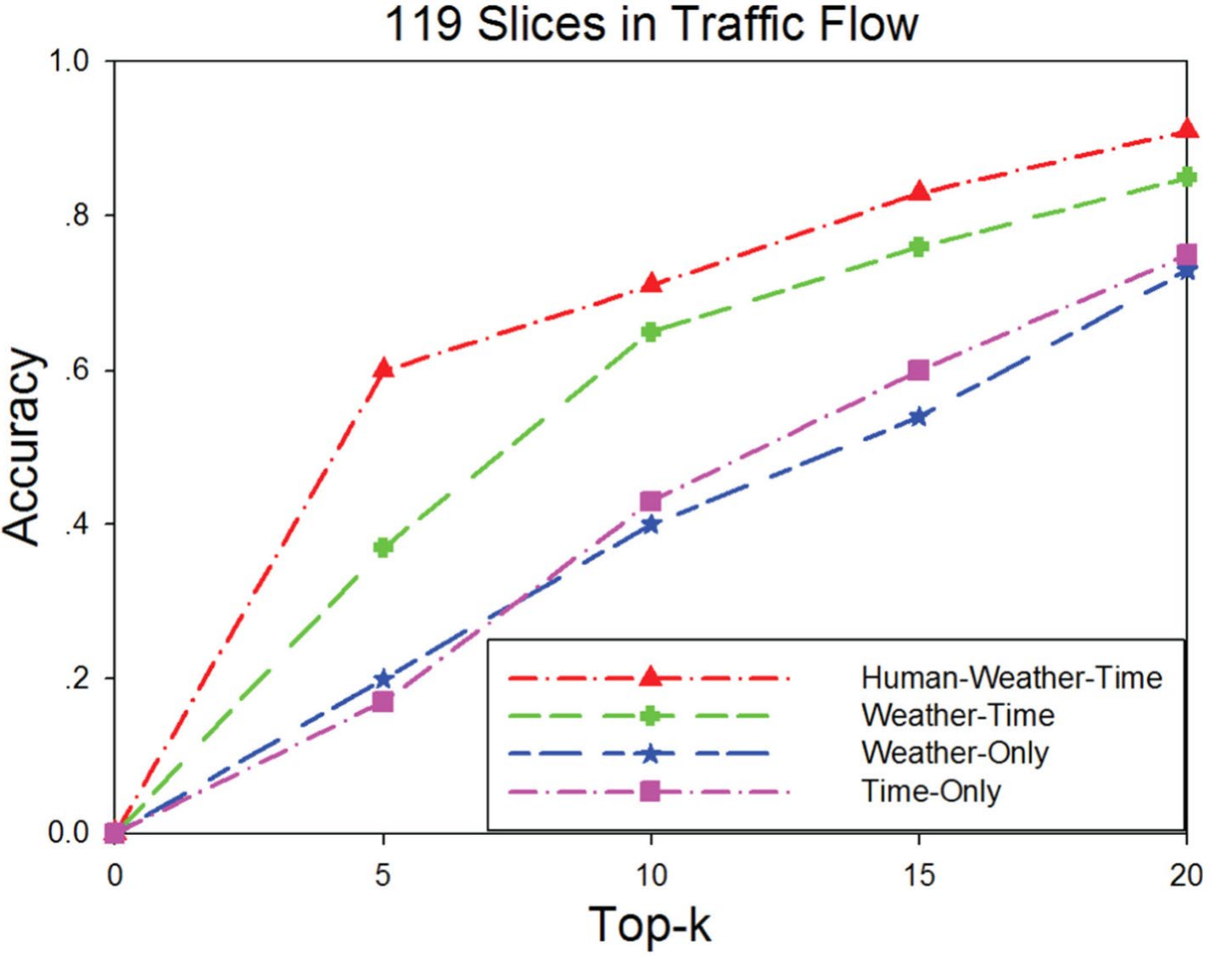}}
	\caption{Prediction accuracy comparison using various Top-k value \cite{wang2018cyber}.}
	\label{Result}
\end{figure}

Similarly, weather, temperature and other information could be fused to generate the transition tensor \cite{wang2018edge}, as shown in Fig. \ref{Transition}. The transition tensor $T$ characterizes the mobility behavior of a user by representing the probability of the user transitioning from one location $P_1$ at a given time $T_1$ to another location $P_2$ at a different time $T_2$. For social factors integration, the mutual influence between humans in terms of mobility pattern is first quantified by the Pearson correlation coefficient (PCC). And the correlation coefficient matrix is normalized in row and obtain $\Lambda$. The social-spatio-temporal tensor is thus attained by the Kronecker product of $T$ and $\Lambda$. To get the largest eigentensor for the prediction result, the eigentensor formula is modified as:
\begin{equation}
X=(1-\alpha)T_{*3}X+\alpha \cdot A
\end{equation}
where tensor operation Einstein production is performed between tensor $T$ and $X$, and $\alpha$ is a hyperparameter that controls the importance of the assigned
probability distributions. And the tensor power method (TPM) is usually introduced to solve this limiting stationary probability distribution problem \cite{8701474, wang2018cyber}. With this framework, user mobility pattern and traffic flow could be predicted with high accuracy. The comparison result between multi-modal prediction and unimodal prediction is shown in Fig. \ref{Result}. As we can see, the Human-Weather-Time model always achieves the highest prediction performance among unimodal models, which implies that using multi-source data greatly improves the prediction accuracy.

\begin{figure}[ht]
	\centerline{\includegraphics[width=20.pc]{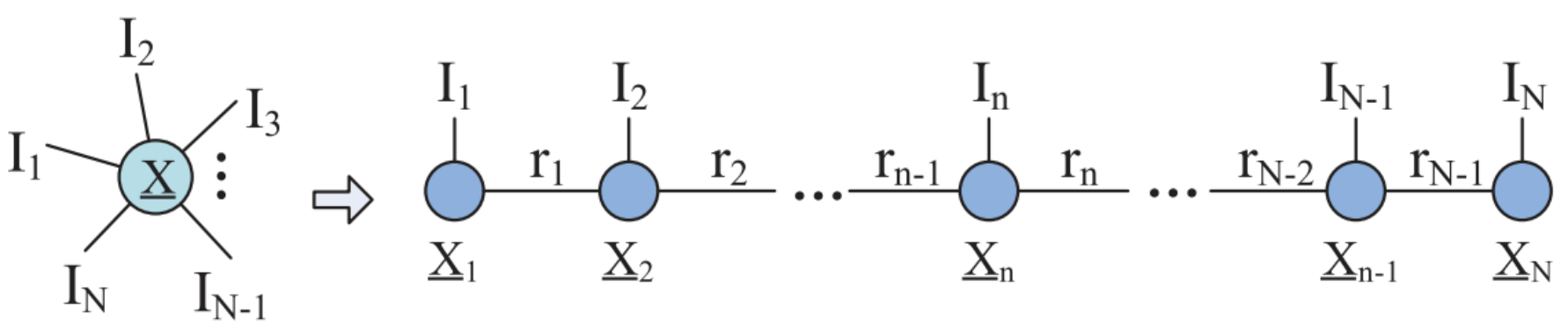}}
	\caption{TT decomposition of an $N$th-order tensor into low-order tensors \cite{9208706}.}
	\label{Decomposition}
\end{figure}

While this multivariate transition tensor method is known for its accurate predictions, its practical implementation is still fraught with challenges stemming from the curse of dimensionality that arises due to the high-order nature of the tensor. On one hand, trying to store and implement high-order dominant eigen decomposition directly based on these original tensors requires loading the entire tensor into memory, which consumes a significant amount of time and memory resources. Consequently, the curse of dimensionality poses a major bottleneck to providing high-quality and rapid services. On the other hand, a single compute node on the cloud or edge/fog devices may not be capable of performing the entire tensor operation, thereby necessitating a solution that can break down each tensor operation into smaller components that can be adapted to run on each node individually.

In order to address this multivariate high-order dominant eigen decomposition problem, several works \cite{9208706, 8449102, liu2022tensor} propose tensor train (TT)-based methods for a scalable decomposition approach which could be implemented on edge devices in a distributed mode, as shown in Fig. \ref{Decomposition}. TT decomposition has been widely used due to its many benefits, such as fewer parameters, distributed storage of TT cores, and the ability to perform TT-based tensor operations. Given an $N$th-order tensor $\underline{X}$, the TT-format should satisfy:
\begin{equation}
\underline{X}=\underline{X_1} \bullet \underline{X_2} \bullet \cdots \underline{X_N} 
\end{equation}
where $\underline{X_n} \in R^{r_{n-1}\times r_n \times I_n}(n=1, \cdots, N; r_0=r_N=1)$ indicates core tensor, $\{r_0, r_1, \cdots,r_N\}$ refers to TT-ranks, and $\bullet$ represents the contraction operation. TT decomposition involves decomposing a higher-order tensor into a sequence of lower-order core tensors. For scalable implementation, TT cores are stored on edge devices, and a tensor train-based higher order dominant Z-eigen decomposition (TT-HODZED) is proposed with Einstein product and Frobenius norm conducted on TT cores.

\subsection{Knowledge-based multi-source data fusion}

\begin{figure*}[ht]
\centering
\includegraphics[width=1\linewidth]{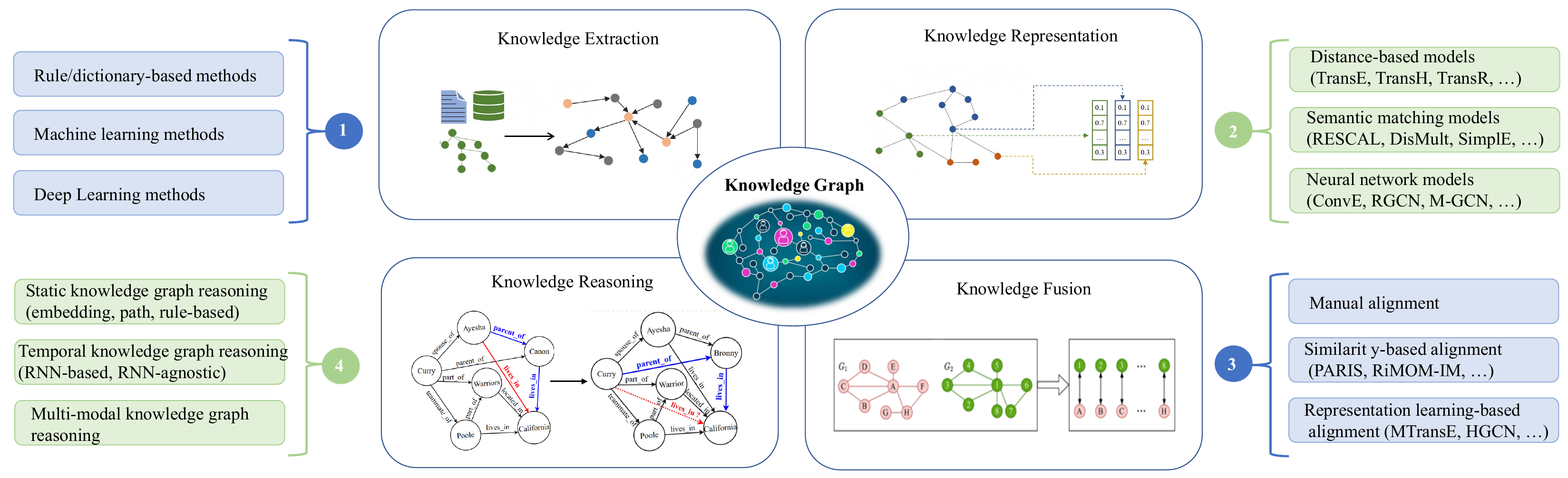}
\caption{An overview of knowledge fusion}
\label{fusion}
\end{figure*}

In addition to the aforementioned data-driven fusion methods, knowledge-driven approaches could achieve fusion at a lower cost. KGs are graph-structured representation of knowledge, encompassing entities (things, concepts, places, etc.), the relationships between these entities, and, in many cases, the attributes of these entities and relationships. KGs aim to model the real-world structure of knowledge in a form that is computable and usable by machines and have been verified to significantly improve the performance of data fusion applications, such as recommendation systems and information retrieval. Specifically, the stages of knowledge fusion include knowledge extraction, knowledge representation, knowledge fusion and knowledge reasoning, as illustrated in Fig. \ref{fusion}.

\subsubsection{Knowledge extraction} 

Knowledge is typically stored and propagated not only in a unstructured formats, but also in a semi-structured or structured formats, such as XML documents, JSON documents and tables. Knowledge extraction is often seen as the first step in knowledge fusion, of which the core is  to extract entities, relationships, attributes, and other knowledge elements from a large amount of heterogeneous data. It can be divided into entity extraction, relation extraction, attribute extraction and event extraction from the perspective of the targets to be extracted. And its implementing methods include rule/dictionary-based methods, machine learning methods and deep learning methods \cite{Yu2020extra, zhuang2023knowledge}. 

\begin{figure}[ht]
\centering
\includegraphics[width=1\linewidth]{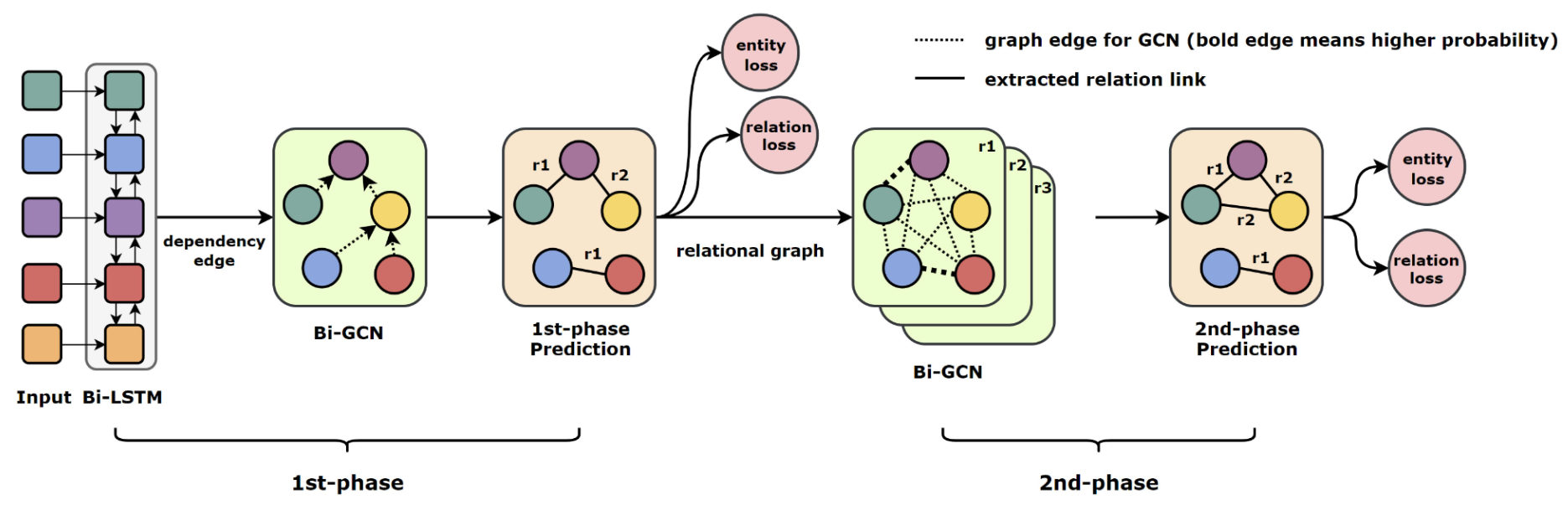}
\caption{Overview of GraphRel with 2nd-phase relation-weighted GCN \cite{fu2019graphrel}.}
\label{extraction}
\end{figure}

Rule/dictionary-based methods use existing rule systems and symbol processing systems to write templates and rules for target objects and then extract named entities, relations, or events by matching them against the text. Segura-Bedmar et al. combine shallow parsing, syntactic simplification and pattern matching to develop a linguistic hybrid rule-based method for the extraction of drug-disease interactions \cite{segura2011linguistic}.  However, this method requires a certain level of expertise and is difficult to cover all element types, resulting in poor portability. Machine learning methods convert knowledge extraction into classification problems, of which the key is the construction of classifiers and feature selection. Pawar et al. design a constrained subsequence kernel and integrate it with SVM classifier for N-ary cross-sentence relations \cite{pawar2020extracting}. Deep learning methods provide a powerful automatic feature extractor, which has been proven that it can achieve excellent performances in knowledge extraction. Fu et al. present an end-to-end relation extraction model named GraphRel based on graph convolutional networks (GCNs) \cite{xu2023generic} to jointly learn named entities and relations, as illustrated in Fig. \ref{extraction}. Considering KGs as undirected graphs, GraphRel utilizes  bi-GCN to aggregate neighbor features and generate updated node features with the following equation:

\begin{equation}
\overrightarrow {h_u}^{l+1}=RELU(\sum_{v\in \overrightarrow{N}(u)}(\overrightarrow {W}^lh_v^l+\overrightarrow {b}^l))
\end{equation}

\begin{equation}
\overleftarrow {h_u}^{l+1}=RELU(\sum_{v\in \overleftarrow{N}(u)}(\overleftarrow {W}^lh_v^l+\overleftarrow {b}^l))
\end{equation}

\begin{equation}
{h_u}^{l+1}=\overrightarrow{h_u}^{l+1} \bigoplus \overleftarrow{h_u}^{l+1} 
\end{equation}

\noindent where ${h_u}^{l}$ denotes the hidden features of node $u$ in the $l$-th layer, $\overrightarrow{N}(u)$ contains all neighboring nodes outgoing from node $u$ and $\overleftarrow{N}(u)$ contains all neighboring nodes incoming from node $u$, both including node $u$. \overrightarrow {W}, \overleftarrow {W}, \overrightarrow {b}, \overleftarrow {b} are learnable convolutional weights and bias.

\subsubsection{Knowledge representation} 

Knowledge representation learning refers to transforming knowledge into low-dimensional vectors to relieve dimensional pressure and capture semantic features. It can project knowledge from different sources into a unified representation space to make a preparation for the subsequent knowledge fusion and knowledge reasoning. 

\begin{figure}[ht]
\centering
\includegraphics[width=1\linewidth]{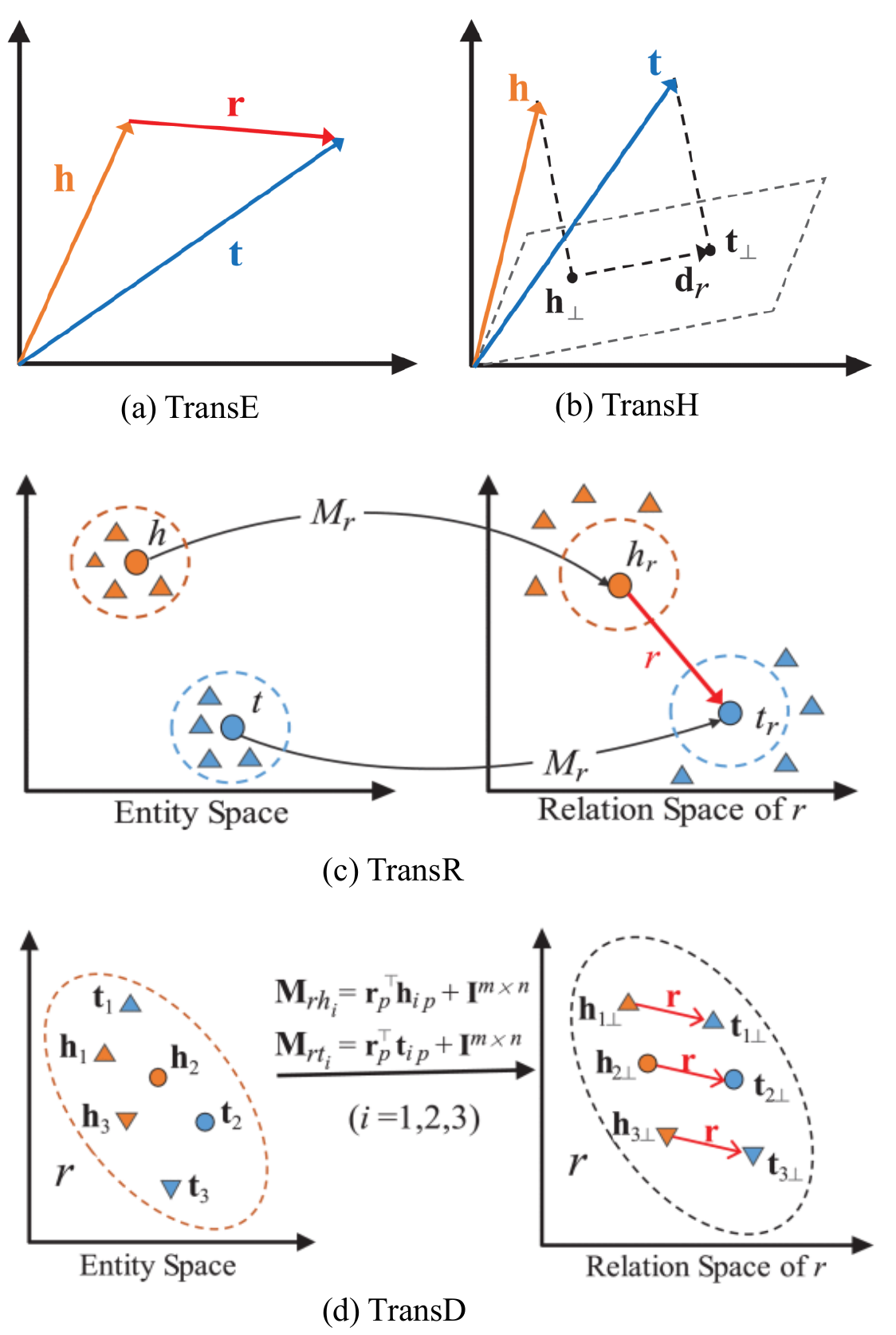}
\caption{Overview of translational models \cite{Chang2017knowledge}.}
\label{translation}
\end{figure}

We classify knowledge presentation learning into three aspects, i.e., distance-based models, semantic matching models, and neural network models \cite{Li2020ASO}. Distance-based models calculate the distance between two entities as scoring functions to measure the rationality of facts. TransE, as one of the most influential models, makes an assumption that a valid triplet $(h,r,t)$ must satisfy a simple translational condition: $ h+r \approx t$ \cite{bordes2013translating}. Therefore, TransE (see Fig.\ref{translation} (a)) minimizes a margin-based ranking criterion to learn embedding as the following:

\begin{equation}
\mathcal L=\sum_{(h,l,t)\in S}\sum_{(h',l,t')\in S'_{(h,l,t)}}[\gamma+d(h+l,t)-d(h'+l,t')]_+
\end{equation}

\noindent where $[x]_+$ represents the positive parts of $x$, $\gamma>0$ is a margin hyperparameter, and $ S'_{(h,l,t)}=\lbrace{(h',l,t)|h'\in E}\rbrace \bigcup \lbrace{(h,l,t')|t'\in E}\rbrace $ is the set of corrupted triplets constructed by replacing either the head $h$ or tail $t$ with a random entity ($h'$ or $t'$). $d(x,y)$ denotes the dissimilarity measure between $x$ and $y$. 

Subsequently, TransH \cite{wang2014knowledge}, TransR \cite{wang2014knowledge}, and TransD \cite{ji2015knowledge}, (see Fig.\ref{translation} (b), (c), (d)) are proposed to overcome the limitations of TransE for better expressive abilities. Semantic matching models compute the semantic similarity-based score functions to measure the possibility of the existence of fact triplets by matching the inclusion relations and the latent semantics of entities. RESCAL \cite{nickel2011three} is proposed as a tensor-based relational learning approach that captures the latent semantics of each entity with vectors and models the pairwise interactions among latent components as a matrix. Other representative works include DistMult \cite{yang2014embedding}, SimplE \cite{kazemi2018simple}, ComplEx \cite{trouillon2016complex}, HolE \cite{nickel2016holographic}, etc. According to the structure of networks, neural network models can be defined based on three types, i.e., traditional neural networks, CNNs and graph neural networks (GNNs). SME first utilizes neural networks to embed multi-relational graphs into a flexible continuous vector space \cite{bordes2014semantic}. ConvE and ConvKB perform 2D convolutions over embeddings \cite{dettmers2018convolutional,nguyen2017novel}. RGCN, M-GCN, and KBGCN aggregate neighborhood information based on GCNs, an evolutionary version of GNN \cite{schlichtkrull2018modeling,wang2019robust,nathani2019learning,wang2019logic}. 

\subsubsection{Knowledge fusion}

Knowledge fusion is also known as ontology alignment, ontology matching, record linkage, entity resolution, and entity alignment \cite{yue2023knowlenet}. Its fundamental problem is to investigate how to integrate descriptive information about the same entities or concepts from multiple sources \cite{zhao2020multi, wang2022multi}. According to the construction of KGs, multi-source knowledge fusion can be divided into two categories: one is knowledge updating for existing KGs, and the other is the fusion of multiple KGs. 

Early entity alignment methods predominantly utilize crowdsourcing techniques to manually label identical entity pairs within the knowledge bases \cite{lehmann2015dbpedia}. These methods are time-consuming and laborious, and struggling to adapt to the growing scale of knowledge bases. Subsequent methods calculate similarity based on certain symbolic features of entities to identify equivalent entity pairs, of which typical models include PARIS \cite{suchanek2011paris} and RiMOM-IM \cite{shao2016rimom}. However, the methods are dependent on manually defined similarity calculation rules.

Currently, representation learning-based entity alignment methods have become mainstream because of their better accuracy and generalizability. The basic components of the methods contain embedding modules that encode KGs using knowledge presentation learning, interaction modules where the embeddings of different KGs are mapped to the same vector space, and alignment modules that align entities based on the distance or similarity between them \cite{Zhang2020entity}. MTransE is the first translation-based model for entity alignment \cite{Chen2016MultilingualKG}, which uses TransE to obtain entity and relation embeddings of three triples $(h,r,t)$ and deploys axis calibration $S_{a1}$, translation vectors $S_{a2}$ and linear transformations $S_{a3}$ to represent cross-lingual transitions ($L_{i}$ and $L_{j}$) in equations \ref{s1}, \ref{s2} and \ref{s3}:

\begin{equation}\label{s1}
\mathcal S_{a1}=||h-h'||+||r-r'||+||t-t'||
\end{equation}

\begin{equation}\label{s2}
\mathcal S_{a2}=||h+{v_{ij}}^e-h'||+||r+{v_{ij}}^r-r'||+||t+{v_{ij}}^e-t'||
\end{equation}

\begin{equation}\label{s3}
\mathcal S_{a3}=||{M_{ij}}^eh-h'||+||{M_{ij}}^rr-r'||+||{M_{ij}}^et-t'||
\end{equation}

\noindent where ${v_{ij}}^e$ represents the entity-dedicated and relation-dedicated translation vectors between $L_{i}$ and $L_{j}$. $M_{ij}^e$ and $M_{ij}^r$ are square matrices of linear transformations on entity and relation vectors from $L_{i}$ to $L_{j}$, respectively. 

Subsequently, many typical models are born, such as IPTransE \cite{Zhu2017IterativeEA}, BootEA \cite{Sun2018BootstrappingEA}, NAEA \cite{Zhu2019NeighborhoodAwareAR}, TransEdge \cite{Sun2019TransEdgeTR}, AKE \cite{Lin2019GuidingCE}, KDCoE \cite{Chen2018CotrainingEO}, etc. Additionally, there are a growing number of works for entity alignment based on GNNs because of their excellent ability to encode the structure information of KGs, such as GCN-Align \cite{Wang2018CrosslingualKG}, HGCN \cite{Wu2019JointlyLE},  KEGCN \cite{Li2019SemisupervisedEA} and AliNet \cite{Sun2019KnowledgeGA}.

\begin{figure}[ht]
\centering
\includegraphics[width=1\linewidth]{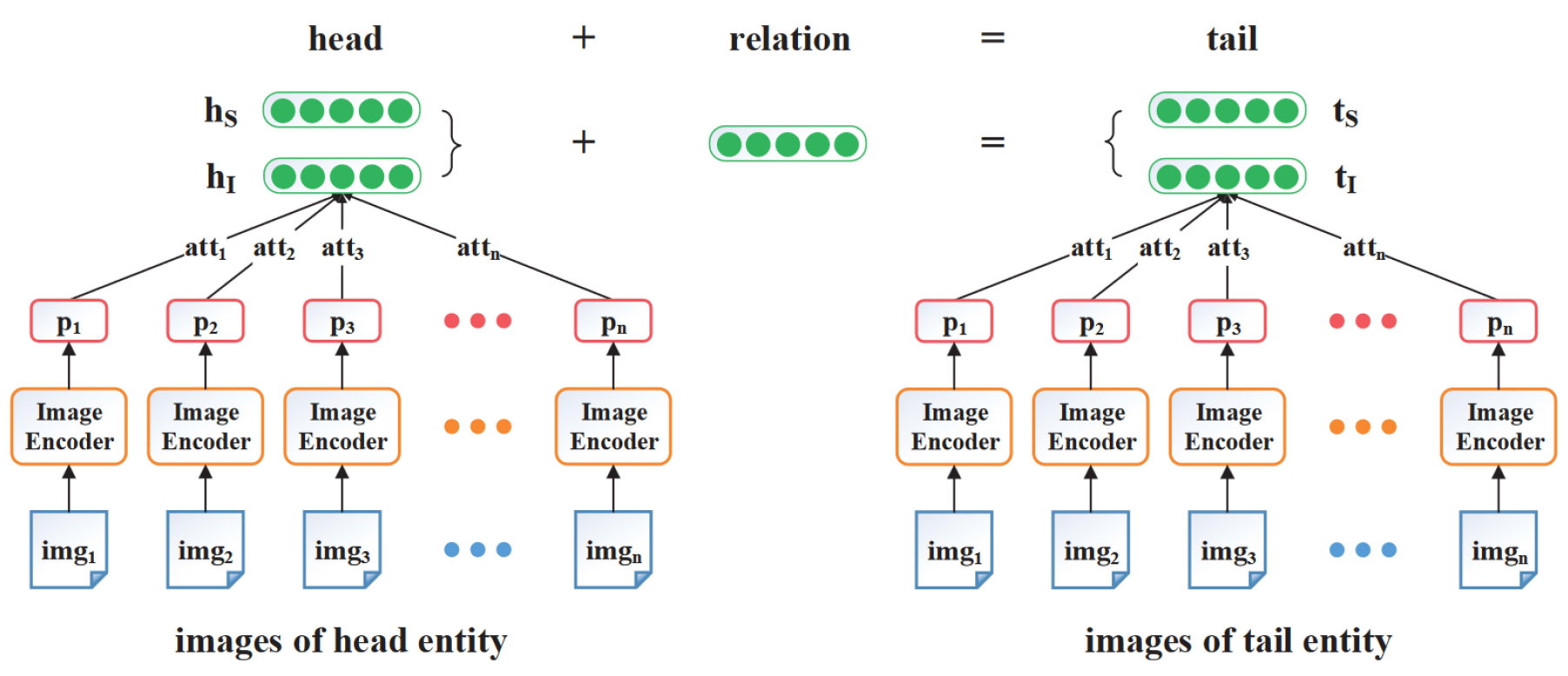}
\caption{ Structure of the IKRL model \cite{Xie2016ImageembodiedKR}.}
\label{IKRL}
\end{figure}

\subsubsection{Knowledge reasoning}
The main target of knowledge reasoning is to deduce new facts from existing facts based on the derived underlying logic rules and thus to make decisions in advance so as to take great advantage of KGs\cite{dong2023knowledge}. According to the category of KGs, knowledge reasoning can be divided into static KG reasoning, temporal KG reasoning and multi-modal KG reasoning. For static KGs, transductive and inductive reasoning are two different reasoning tasks. Transductive reasoning requires that entities and relations in the queried facts are all seen in the given KGs, while the candidates may be beyond the given graphs within the inductive reasoning. For temporal KGs, in terms of whether the occurrence time of queried facts are within the time or after the time covered by the given KGs, reasoning can be defined as interpolation and extrapolation respectively. 

\begin{figure*}
\centering
\includegraphics[width=0.9\linewidth]{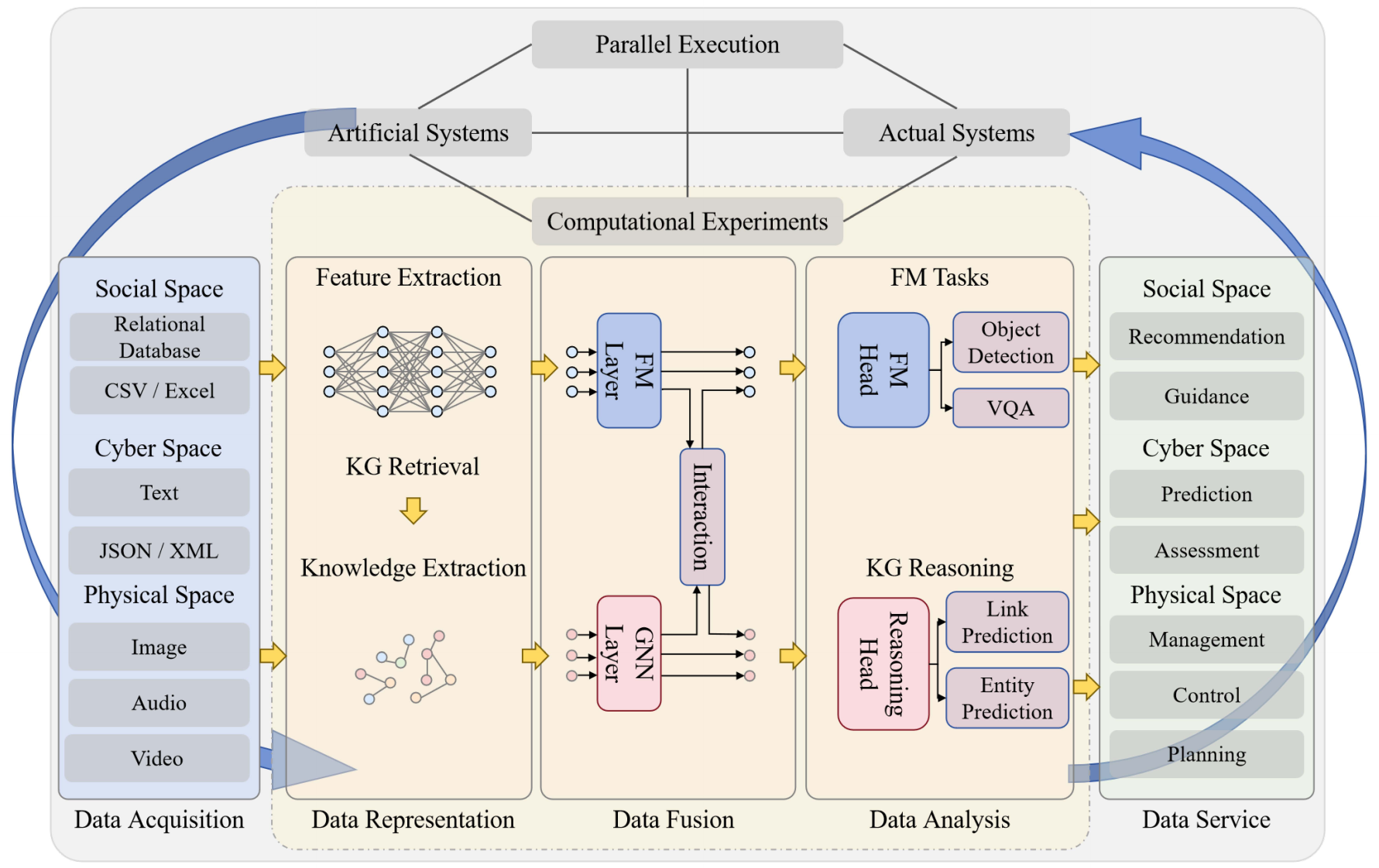}
\caption{Our proposed synergized multi-source data fusion framework for CPSS. The framework consists of 5 layers, the first layer is data acquisition layer, the second layer is data representation layer, the third layer is data fusion layer, the fourth layer is data analysis layer, the fifth layer is data service layer.}
\label{Framework}
\end{figure*}

Static KG reasoning can be classified into embedding-based models, path-based models and rule-based models. Embedding-based models include translation models such as TransE, TransH and TransD \cite{bordes2013translating,wang2014knowledge,ji2015knowledge}, tensor compositional models like RESCAL, DistMult, and ComplEX \cite{nickel2011three,yang2014embedding,trouillon2016complex}, and neural network models such as SME, ConvE and RGCN \cite{bordes2014semantic, dettmers2018convolutional, schlichtkrull2018modeling}. Path-based models use advanced technologies such as deep reinforcement learning, random walk and the path-ranking algorithm (PRA) to mine the explicit logic knowledge underlying the path between heads and tails, including ProPPR, RNNPRA, DeepPath, etc \cite{Gardner2014IncorporatingVS,Neelakantan2015CompositionalVS,Xiong2017DeepPathAR}. Rule-based models make full use of symbolic features to extract logical rules: $B \rightarrow A$, where $A$ presents a fact and $B$ denotes a set of facts, and thus inject them into embedding for better reasoning performances, including AMIE, KALE, RUGE, etc \cite{Galrraga2013AMIEAR, Guo2016JointlyEK, Guo2017KnowledgeGE}. Temporal KGs add an extra dimension of time information to static KGs. RNNs are well-suited for capturing temporal dynamics, so RNNs and their variants are widely applied in temporal KG reasoning. Know-Evolve learns non-linearly evolving entity representations over time based on an RNN-based deep evolutionary knowledge network \cite{Trivedi2017KnowEvolveDT}. TTransE leverages long short-term memory (LSTM) networks to encode time evolution into translations based on TransE \cite{GarcaDurn2018LearningSE}. TeMP is proposed to integrate structure information across time by using gated recurrent units (GRU) \cite{Wu2020TeMPTM}. Additionally, some models incorporate time information without RNN frameworks and result in excellent performances. GyGNet first introduces the copy mechanism into the temporal reasoning model \cite{Zhu2020LearningFH}. DA-Net is inspired by human decision-making processes and models dynamic distributions of historical repetitive facts via distributed attention \cite{Liu2022DANetDA}. Representatives of multi-modal KG reasoning include CKE, DKRL, IKRL, etc \cite{Zhang2016CollaborativeKB, Xie2016RepresentationLO, Xie2016ImageembodiedKR}. They generally fuse the multi-modal auxiliary features into embedding-based reasoning models to deduce queried facts. For example, inspired by TransE, IKRL (see Fig. \ref{IKRL}) defines the overall energy function as follows: 

\begin{equation}
\mathcal E(h,r,t)=E_{SS}+E_{SI}+E_{IS}+E_{II}
\end{equation}

\noindent where $E_{SS}$ and $E_{II}$ are the same energy function as TransE based on structure-based and image-based representations respectively, and $E_{SI}$ and $E_{IS}$ assure that the two types of representations are learned into the same vector space.

\begin{table*}[ht]
\centering
\caption{Production pipeline of parallel weaving}
\label{table:line}
\begin{tabular}{ll}
\hline
Production stage & Functions\\ \hline
Pre-production design & Customization, collaborative design\\
Supply chain optimization & Procurement optimization, real-time material tracking \\
Smart manufacturing & Human-machine collaboration, flexible production lines, predictive maintenance, and AR demonstration \\
Quality assurance & Real-time monitoring, closed-loop feedback \\
Post-production distribution & Customized packaging, sustainable logistics\\\hline

\end{tabular}
\end{table*}

\section{A synergized CPSS multi-source data fusion framework for industrial metaverses}

Considering the strengths of deep learning and KGs in scalability and parallel computation, we combine both approaches, proposing a synergized multi-source data fusion framework. This integration capitalizes on the complementary strengths of both, addressing the shortcomings of deep learning in terms of explainability and fact fabrication, as well as overcoming the incompleteness and the challenges of construction and maintenance inherent in KGs \cite{pan2023unifying}. However, the nature of the knowledge differs between the two: in FMs, it is typically implicit and lacks a defined structure, whereas in KGs, it is explicit and well-structured. Thus, aligning the knowledge from unstructured data with that of KGs is essential for representing them in a cohesive and unified manner.

Inspired by exsiting work \cite{wang2021kepler, ke2021jointgt, yasunaga2022deep}, we propose a synergized multi-modal data fusion framework by integrating KGs with FM for industrial metaverses, as depicted in Fig. \ref{Framework}. Data representation, data fusion, and data analysis are performed to achieve specific tasks such as recommendation, prediction, planning, etc. Specifically, we construct paired alignments (text/image/audio/video, local KG) by selecting segments from multi-modal sources and deriving associated subgraphs from the KG. The KG's structured knowledge serves to anchor the data, and conversely, the data enriches the KG with a dense context for logical deductions. Our goal is to collaboratively pretrain a FM using these aligned pairs. This framework employs a cross-modal encoder to mediate this two-way information flow, where each layer of the model encodes the raw data with an FM and the KG with a GNN, and fuses the two with a bidirectional modality interaction module, resulting in integrated representations. During the pretraining phase, we unify specific tasks related to both the KG and the FM. This combined objective prompts the various data types to reciprocally inform and enhance each other, enabling the model to make more precise predictions.

\section{Application in Parallel Weaving}

\begin{figure*}[ht]
\centering
\includegraphics[width=.75\linewidth]{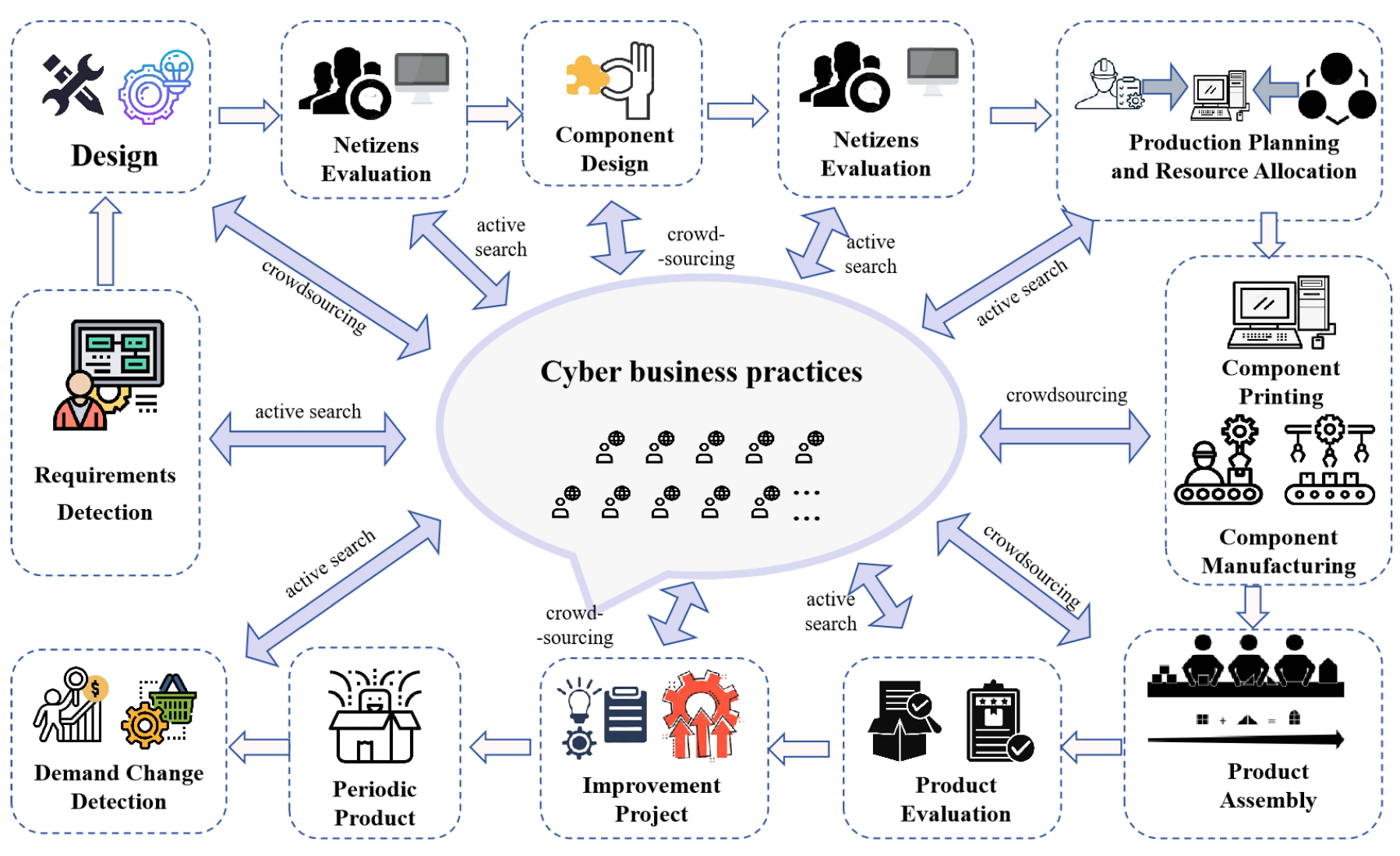}
\caption{Overview of social manufacturing.}
\label{SM}
\end{figure*}

\begin{figure}[ht]
\centering
\includegraphics[width=\linewidth]{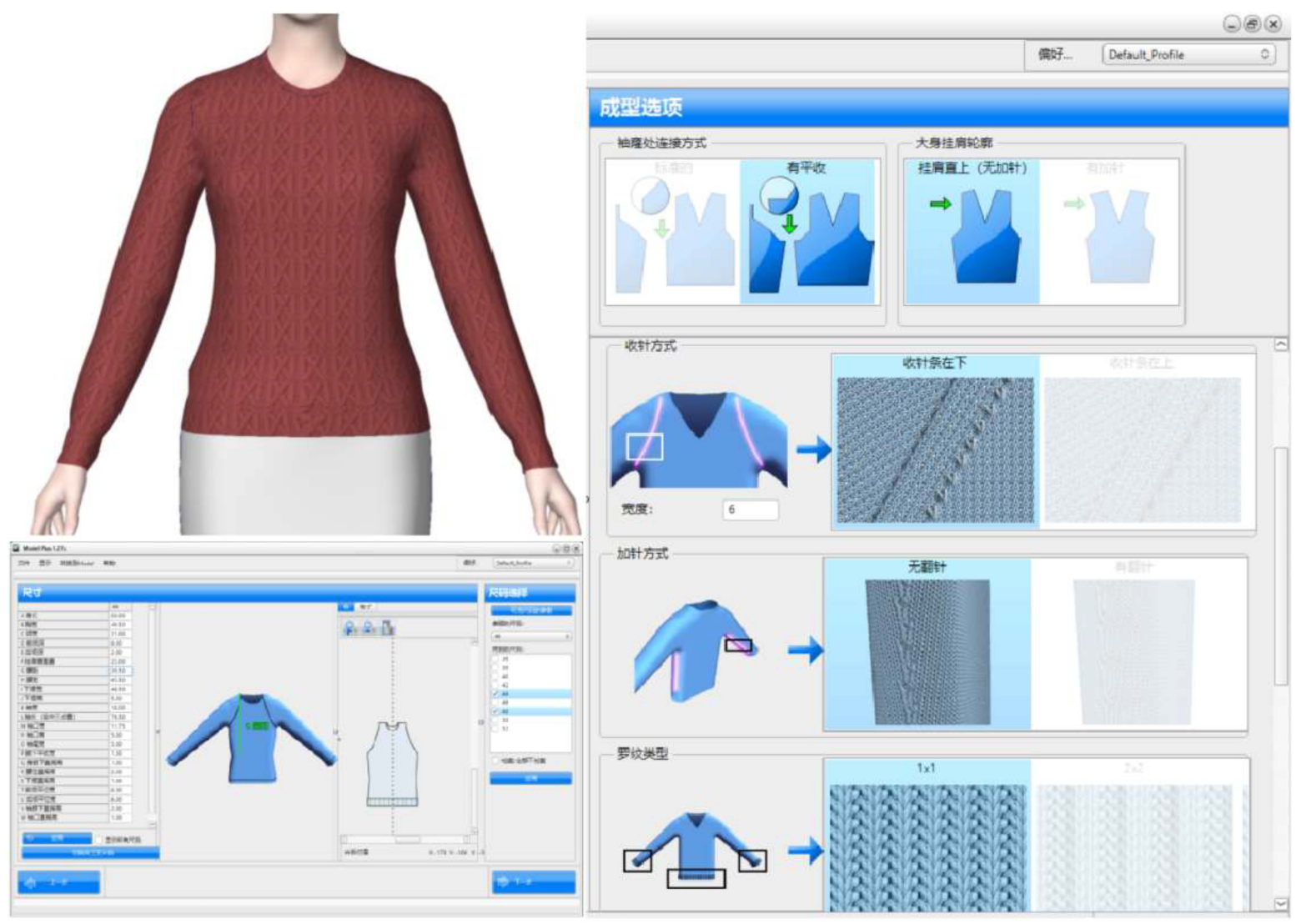}
\caption{Customized Design and production planning.}
\label{Design}
\end{figure}

To realize the management and control of complex systems, Fei-Yue Wang first proposed the Artificial systems, Computational experiments, and Parallel execution (ACP) method \cite{wang2004parallel, wang2016steps, cao2023parameter} in 2004. Essentially, the implementation of ACP could be divided into three steps: (1) Constructing artificial systems that correspond to real complex systems; (2) Utilizing computational experiments to train, predict, and evaluate complex systems; (3) Realizing effective control and management over the complex systems through parallel execution.

Based on parallel theory, parallel manufacturing was first proposed in 2018 \cite{wangfeiyue2018Parallel, li2018Parallel, li2021distributed, yang2022parallel, yang2022defact} to realize quick response to changing demands, reduce downtime, minimize waste, improve product quality, and create a more agile and competitive manufacturing environment. In parallel manufacturing, multi-source heterogeneous data is collected from actual physical, cyber, and social space, which includes unstructured data such as images, audio, text, and video, semi-structured data like XML and JSON, and structured data from databases, CSV, etc. With the data, an industrial metaverse is built, containing digital replicas of entities in the entire manufacturing CPSS, and including stakeholders, machinery, and components. The industrial metaverse serves as the artificial manufacturing system corresponding to the actual one, providing a 3D virtual environment where users can interact. Within the industrial metaverse, the synergized multi-source data fusion framework is deployed to perform computational experiments for perception, prediction, and planning tasks, thus achieving intelligent management and control, etc. For parallel execution, we compare the differences between the actual feedback and artificial results, and could accordingly improve the industrial metaverse construction and data fusion model. As these three steps iterate, the optimization improves the quality of industrial metaverses and the performance of data fusion model. During this process, the power of social networks, cloud computing, and advanced technologies to foster collaboration between producers, consumers, and stakeholders across the manufacturing ecosystem are fully leveraged in a social manufacturing pattern, as shown in Fig. \ref{SM}.

Parallel manufacturing has already been applied in Ningbo Cixing Co., Ltd. for knitted apparel industry, especially sweater manufacturing, as shown in Fig. \ref{Design}. The production pipeline in parallel manufacturing reflects the convergence of human intuition and skills with the precision and efficiency of smart technologies. This integration prioritizes customization, sustainability, and human-machine collaboration. We detail the sweater manufacturing pipeline as follows, and list it in Table \ref{table:line}.


\begin{table*}[ht]
  \centering
  \caption{Challenges and future directions for multi-source data fusion in CPSS}\label{table: CF}
  {
    \begin{tabularx}{1.0\linewidth}
    { p{1.8cm} p{4.5cm} p{5.4cm} p{4.5cm} }
      \hline
      Directions & Current shortcomings& Future challenges&Challenges-solving benefits\\ \hline
      Generality & The current model design is based on prior knowledge of input types. & Development of robust input-agnostic model design techniques.& Build a general model for a variety of inputs efficiently.\\ \hline
      Timeliness &  The collection of multi-source data greatly challenges computational efficiency due to the increased data volume.& Exploration of the distributed processing frameworks based on edge computing and parallel computing. & Distribute data fusion processing to edge devices for real-time, large-scale data fusion while maintaining data integrity. \\ \hline
      Security & Data exchange and sharing between devices pose risks of sensitive information disclosure. & Design of efficient and reliable consensus and incentive mechanisms based on federated learning, blockchain and smart contracts. & Enable secure data exchange and sharing while ensuring efficiency. \\
      \hline
    \end{tabularx}
  }
\end{table*}

\begin{itemize}

\item Pre-production design:
\begin{itemize}
\item Customization: Manufacturers realize fashion trend forecasting and unique customer preferences identification to design customized sweaters.

\item Collaborative design: Humans and generative AI tools collaboratively work on product designs, incorporating the best of human creativity and machine precision.
\end{itemize}

\item Supply chain optimization:
\begin{itemize}
\item Procurement optimization: It refers to the comprehensive approach of achieving maximum value creation through the integration of people, processes, and technology.

\item Real-time material tracking: With data gathered from sensors and IoT devices, real-time tracking of materials and components is conducted to ensure timely availability.

\end{itemize}
\item Smart manufacturing:
\begin{itemize}
\item Human-machine collaboration: `Cobots' (collaborative robots) work alongside humans, assisting in tasks that require precision or repeatability, ensuring both efficiency and safety.

\item Flexible production lines: Adaptive machinery can easily switch between different products or product variants, allowing for greater customization.

\item Predictive maintenance: Machinery health is constantly monitored, predicting when maintenance is due, allowing for proactive measures to minimize downtime.

\item Augmented Reality (AR) demonstration: AR tools guide workers through complex assembly processes, maintenance tasks, or quality checks.
\end{itemize}
\item Quality assurance:
\begin{itemize}
\item Real-time monitoring: Sensors and cameras constantly monitor the production process, catching defects or inconsistencies in real-time.

\item Close-loop feedback: Actual defect data from the quality checks is fed into the system to constantly refine and improve the production and monitoring process.
\end{itemize}
\item Post-production distribution:
\begin{itemize}
\item Customized packaging: Products are packaged according to individual specifications or customer requirements.
\item Sustainable logistics: Logistics are optimized for the quickest, most fuel-efficient routes, reducing carbon footprints.

\end{itemize}
\end{itemize}

\section{Challenges and Future Directions}

To gain comprehensive insights and effective decision policies, multi-source data fusion in CPSS has already become the mainstream research direction. However, there still exist several challenges that need to be addressed. In this section, we point out three major challenges and corresponding future research directions, as shown in Table \ref{table: CF}.

\textbf{The generality of models to various inputs.} Due to the high diversity of inputs in multi-source data fusion, prior knowledge of input types is necessary to design appropriate model structures in current data fusion methods. For instance, when dealing with a single image as input, a 2D convolutional structure can be utilized to extract relevant features \cite{jogin2018feature}. On the other hand, when the input consists of text, RNNs or Transformer models become more suitable choices \cite{tabassum2020survey}. Hence, for multi-modal inputs, it is crucial to develop dedicated feature extraction modules for each modality input and align them accordingly. Furthermore, we must adapt our architecture whenever there are changes in the number or type of input modalities. This is quite cumbersome, laborious, and time-consuming \cite{jaegle2021perceiver}. Future research will focus on developing robust input-agnostic model design techniques that reduce architectural assumptions about the relationships between inputs to efficiently build a general model for a variety of inputs.

\textbf{Timeliness of data processing.} By combining data from multiple sources, we can gain a comprehensive understanding of various systems for decision-making, whereby widespread applications have emerged in various fields such as intelligent transportation \cite{yinglei2022smart}, smart manufacturing \cite{liu2022intelligent}, and smart agriculture \cite{li2020acquisition}. However, the collection of multi-source data greatly increases the amount of data, which poses a great challenge to computational efficiency. Therefore, appropriate data compression techniques are adopted to reduce the data volume and thus achieve the real-time processing of data fusion. For example, to handle massive amounts of data in training or inferring processes, model compression \cite{choudhary2020comprehensive} is currently employed for deep learning methods. Similarly, dimensionality reduction \cite{kuang2014tensor} is utilized for tensor fusion methods. However, these methods may also lead to the loss of valuable information and the impairment of model precision. Therefore, drawing inspiration from edge computing and parallel computing, we expect to explore distributed processing frameworks in the future, which can distribute the processing burden of data fusion to multiple edge devices to handle large-scale data fusion in real time while ensuring data integrity.

\textbf{Data security and privacy protection}. Typically, data fusion involves the exchange and sharing of data between devices, which can pose a potential risk for the disclosure of sensitive information, such as personal information and business operational data. Therefore, privacy protection \cite{feng2020privacy} and security of the fused data \cite{al2023privacy} should be given enough attention. Currently, some efforts have already been made. For example, homomorphic encryption is incorporated into the Lanczos-based tensor decomposition to enable privacy-preserving and effective computations \cite{feng2018privacy, feng2019practical}. RNN, tensor, and differential privacy are integrated to deal with both heterogeneous IoT data and privacy issues \cite{feng2023tensor}. Some studies apply Laplace mechanism-based differential privacy to crowdsourcing systems for the protection of sensitive information of workers \cite{zhang2022task, zhang2023effective}. However, these methods require a reliable control center, and in the event of an attack on this center, the entire system may face catastrophic collapse. Federated learning \cite{ZHOU2023102182, diaz2023connecting, qi2023fl, yang2023tensor}, blockchain \cite{yuan2018blockchain,Xu2023blockchain,Lang2022blockchain} and smart contracts \cite{wang2019blockchain,Leng2023Blockchain,wang2018overview} have the potential to become powerful tools to address the issue. However, they may give rise to additional challenges. For example, inappropriate consensus mechanisms and incentive mechanisms not only hinder the efficiency of data collection, processing, and storage but also result in the monopolization of decision-making authority. Therefore, how to design efficient and reliable consensus and incentive mechanisms, based on federated learning, blockchain and smart contracts, for enabling secure data exchange and sharing while ensuring efficiency, is an important research direction for the future.

    


Moreover, other challenges such as data quality and attack defense, should also be carefully investigated to pave the way for more extensive data fusion applications.

\section{Conclusion}
In this paper, we introduce a hierarchical operational framework for industrial metaverses, explaining its internal operational mechanisms using multi-source data fusion techniques. Subsequently, we offer an extensive overview of the multi-source data fusion, which serves as the fundamental building block for industrial metaverses. Moreover, to address the limitations of data-driven and knowledge-driven approaches, we present a synergized CPSS multi-source data fusion framework tailored for industrial metaverses, enhancing the performance of perception, prediction, and planning. Furthermore, we demonstrate the application of this proposed architecture in parallel sweater manufacturing, outlining how it interacts with real manufacturing systems and showcasing collaborative design, customization, and remote interaction to validate its efficacy. We also offer a perspective on current challenges and future directions of multi-source data fusion. Throughout the paper, we aim to shed light on the significance and potential of multi-source data fusion in manufacturing CPSS, and encourage further research in this field.

\section*{Acknowledgment}
This work was supported by Prediction and Guidance Effect of Social Media on Traffic Congestion and Its Derivative Events, National Natural Science Foundation of China (62173329).

\bibliographystyle{IEEEtran}
\bibliography{IFRef}

\begin{thebibliography}{100}
\providecommand{\url}[1]{#1}
\csname url@samestyle\endcsname
\providecommand{\newblock}{\relax}
\providecommand{\bibinfo}[2]{#2}
\providecommand{\BIBentrySTDinterwordspacing}{\spaceskip=0pt\relax}
\providecommand{\BIBentryALTinterwordstretchfactor}{4}
\providecommand{\BIBentryALTinterwordspacing}{\spaceskip=\fontdimen2\font plus
\BIBentryALTinterwordstretchfactor\fontdimen3\font minus \fontdimen4\font\relax}
\providecommand{\BIBforeignlanguage}[2]{{%
\expandafter\ifx\csname l@#1\endcsname\relax
\typeout{** WARNING: IEEEtran.bst: No hyphenation pattern has been}%
\typeout{** loaded for the language `#1'. Using the pattern for}%
\typeout{** the default language instead.}%
\else
\language=\csname l@#1\endcsname
\fi
#2}}
\providecommand{\BIBdecl}{\relax}
\BIBdecl

\bibitem{wang2023guest}
F.-Y. Wang, Y.~Tang, and P.~J. Werbos, ``Guest editorial: cyber--physical--social intelligence: toward metaverse-based smart societies of {6I} and {6S},'' \emph{IEEE Trans. Syst. Man Cybern. Syst.}, vol.~53, no.~4, pp. 2018--2024, 2023.

\bibitem{chen2023metaverse}
P.-K. Chen, Y.~Ye, and X.~Huang, ``The metaverse in supply chain knowledge sharing and resilience contexts: An empirical investigation of factors affecting adoption and acceptance,'' \emph{J. Innov. Knowl.}, vol.~8, no.~4, p. 100446, 2023.

\bibitem{lee2022integrated}
J.~Lee and P.~Kundu, ``Integrated cyber-physical systems and industrial metaverse for remote manufacturing,'' \emph{Manuf. Lett.}, vol.~34, pp. 12--15, 2022.

\bibitem{5552591}
F.-Y. Wang, ``The emergence of intelligent enterprises: From {CPS} to {CPSS},'' \emph{IEEE Intell. Syst.}, vol.~25, no.~4, pp. 85--88, 2010.

\bibitem{yilma2021systemic}
B.~A. Yilma, H.~Panetto, and Y.~Naudet, ``Systemic formalisation of cyber-physical-social system (cpss): A systematic literature review,'' \emph{Comput. Ind.}, vol. 129, p. 103458, 2021.

\bibitem{tao2018digital}
F.~Tao, H.~Zhang, A.~Liu, and A.~Y. Nee, ``Digital twin in industry: State-of-the-art,'' \emph{IEEE Trans. Industr. Inform.}, vol.~15, no.~4, pp. 2405--2415, 2018.

\bibitem{liu2021review}
M.~Liu, S.~Fang, H.~Dong, and C.~Xu, ``Review of digital twin about concepts, technologies, and industrial applications,'' \emph{J. Manuf. Syst.}, vol.~58, pp. 346--361, 2021.

\bibitem{yao2022enhancing}
X.~Yao, N.~Ma, J.~Zhang, K.~Wang, E.~Yang, and M.~Faccio, ``Enhancing wisdom manufacturing as industrial metaverse for industry and society 5.0,'' \emph{J. Intell. Manuf.}, pp. 1--21, 2022.

\bibitem{wang2023steps}
X.~Wang, J.~Yang, Y.~Wang, Q.~Miao, F.-Y. Wang, A.~Zhao, J.-L. Deng, L.~Li, X.~Na, and L.~Vlacic, ``Steps toward industry 5.0: Building “{6S}” parallel industries with cyber-physical-social intelligence,'' \emph{IEEE/CAA J. of Autom. Sin.}, vol.~10, no.~8, pp. 1692--1703, 2023.

\bibitem{wang2018societies}
F.-Y. Wang, Y.~Yuan, X.~Wang, and R.~Qin, ``Societies 5.0: A new paradigm for computational social systems research,'' \emph{IEEE Trans. Comput. Soc. Syst.}, vol.~5, no.~1, pp. 2--8, 2018.

\bibitem{wang2019data}
P.~Wang, L.~T. Yang, J.~Li, J.~Chen, and S.~Hu, ``Data fusion in cyber-physical-social systems: State-of-the-art and perspectives,'' \emph{Inf. Fusion}, vol.~51, pp. 42--57, 2019.

\bibitem{meng2020survey}
T.~Meng, X.~Jing, Z.~Yan, and W.~Pedrycz, ``A survey on machine learning for data fusion,'' \emph{Inf. Fusion}, vol.~57, pp. 115--129, 2020.

\bibitem{zhang2021tensor}
S.~Zhang, L.~T. Yang, J.~Feng, W.~Wei, Z.~Cui, X.~Xie, and P.~Yan, ``A tensor-network-based big data fusion framework for cyber-physical-social systems ({CPSS}),'' \emph{Inf. Fusion}, vol.~76, pp. 337--354, 2021.

\bibitem{yang2023representation}
J.~Yang, L.~T. Yang, H.~Wang, Y.~Gao, Y.~Zhao, X.~Xie, and Y.~Lu, ``Representation learning for knowledge fusion and reasoning in cyber-physical-social systems: Survey and perspectives,'' \emph{Inf. Fusion}, vol.~90, pp. 59--73, 2023.

\bibitem{zhang2021multi}
P.~Zhang, T.~Li, G.~Wang, C.~Luo, H.~Chen, J.~Zhang, D.~Wang, and Z.~Yu, ``Multi-source information fusion based on rough set theory: A review,'' \emph{Inf. Fusion}, vol.~68, pp. 85--117, 2021.

\bibitem{khaleghi2013multisensor}
B.~Khaleghi, A.~Khamis, F.~O. Karray \emph{et~al.}, ``Multisensor data fusion: A review of the state-of-the-art,'' \emph{Inf. Fusion}, vol.~14, no.~1, pp. 28--44, 2013.

\bibitem{chai2023heterogeneous}
L.~Chai, W.~Yi, R.~Hoseinnezhad, and L.~Kong, ``Heterogeneous multi-sensor fusion for {PHD} filter in decentralized sensor networks,'' \emph{Inf. Fusion}, vol. 100, p. 101956, 2023.

\bibitem{qu2023qnmf}
Z.~Qu, Y.~Li, and P.~Tiwari, ``{QNMF}: A quantum neural network based multimodal fusion system for intelligent diagnosis,'' \emph{Inf. Fusion}, vol. 100, p. 101913, 2023.

\bibitem{nie2020multi}
R.~Nie, J.~Cao, D.~Zhou, and W.~Qian, ``Multi-source information exchange encoding with pcnn for medical image fusion,'' \emph{IEEE Trans. Circuits Syst. Video Technol.}, vol.~31, no.~3, pp. 986--1000, 2020.

\bibitem{tao2022multi}
X.~Tao and J.~D. Velasquez, ``Multi-source information fusion for smart health with artificial intelligence,'' pp. 93--95, 2022.

\bibitem{zhang2023multi}
P.~Zhang, T.~Li, G.~Wang, D.~Wang, P.~Lai, and F.~Zhang, ``A multi-source information fusion model for outlier detection,'' \emph{Inf. Fusion}, vol.~93, pp. 192--208, 2023.

\bibitem{cai2014multi}
B.~Cai, Y.~Liu, Q.~Fan, Y.~Zhang, Z.~Liu, S.~Yu, and R.~Ji, ``Multi-source information fusion based fault diagnosis of ground-source heat pump using bayesian network,'' \emph{Appl. Energy}, vol. 114, pp. 1--9, 2014.

\bibitem{li2018physics}
H.~Li, H.-Z. Huang, Y.-F. Li, J.~Zhou, and J.~Mi, ``Physics of failure-based reliability prediction of turbine blades using multi-source information fusion,'' \emph{Appl. Soft Comput.}, vol.~72, pp. 624--635, 2018.

\bibitem{guo2018mobile}
Y.~Guo, C.~Yin, M.~Li, X.~Ren, and P.~Liu, ``Mobile e-commerce recommendation system based on multi-source information fusion for sustainable e-business,'' \emph{Sustainability}, vol.~10, no.~1, p. 147, 2018.

\bibitem{cheng2019friend}
S.~Cheng, B.~Zhang, G.~Zou, M.~Huang, and Z.~Zhang, ``Friend recommendation in social networks based on multi-source information fusion,'' \emph{Int. J. Mach. Learn. Cybern.}, vol.~10, pp. 1003--1024, 2019.

\bibitem{smirnov2019knowledge}
A.~Smirnov and T.~Levashova, ``Knowledge fusion patterns: A survey,'' \emph{Inf. Fusion}, vol.~52, pp. 31--40, 2019.

\bibitem{zhao2020multi}
X.~Zhao, Y.~Jia, A.~Li, R.~Jiang, and Y.~Song, ``Multi-source knowledge fusion: A survey,'' \emph{World Wide Web}, vol.~23, pp. 2567--2592, 2020.

\bibitem{10123038}
P.~Xu, X.~Zhu, and D.~A. Clifton, ``Multimodal learning with {Transformers}: A survey,'' \emph{IEEE Trans. Pattern Anal. Mach. Intell.}, pp. 1--20, 2023.

\bibitem{han2023survey}
X.~Han, Y.-T. Wang, J.-L. Feng, C.~Deng, Z.-H. Chen, Y.-A. Huang, H.~Su, L.~Hu, and P.-W. Hu, ``A survey of {Transformer}-based multimodal pre-trained modals,'' \emph{Neurocomputing}, vol. 515, pp. 89--106, 2023.

\bibitem{guo2019deep}
W.~Guo, J.~Wang, and S.~Wang, ``Deep multimodal representation learning: A survey,'' \emph{IEEE Access}, vol.~7, pp. 63\,373--63\,394, 2019.

\bibitem{wang2024human}
B.~Wang, H.~Zhou, X.~Li, G.~Yang, P.~Zheng, C.~Song, Y.~Yuan, T.~Wuest, H.~Yang, and L.~Wang, ``Human digital twin in the context of industry 5.0,'' \emph{Robot. Comput. Integr. Manuf.}, vol.~85, p. 102626, 2024.

\bibitem{zhang2023towards}
C.~Zhang, Z.~Wang, G.~Zhou, F.~Chang, D.~Ma, Y.~Jing, W.~Cheng, K.~Ding, and D.~Zhao, ``Towards new-generation human-centric smart manufacturing in industry 5.0: A systematic review,'' \emph{Adv. Eng. Inform.}, vol.~57, p. 102121, 2023.

\bibitem{wang2022metaverse}
F.-Y. Wang, ``The metaverse of mind: Perspectives on {DeSci} for {DeEco} and {DeSoc},'' \emph{IEEE/CAA J. of Autom. Sin.}, vol.~9, no.~12, pp. 2043--2046, 2022.

\bibitem{gadekallu2023blockchain}
T.~R. Gadekallu, W.~Wang, G.~Yenduri, P.~Ranaweera, Q.-V. Pham, D.~B. da~Costa, M.~Liyanage \emph{et~al.}, ``Blockchain for the metaverse: A review,'' \emph{Future Gener. Comput. Syst.}, vol. 143, pp. 401--419, 2023.

\bibitem{wang2022dao}
{F.-Y. Wang}, ``The {DAO} to metacontrol for metasystems in metaverses: The system of parallel control systems for knowledge automation and control intelligence in {CPSS},'' \emph{IEEE/CAA J. of Autom. Sin.}, vol.~9, no.~11, pp. 1899--1908, 2022.

\bibitem{wang2022framework}
J.~Wang, Y.~Tian, Y.~Wang, J.~Yang, X.~Wang, S.~Wang, and O.~Kwan, ``A framework and operational procedures for metaverses-based industrial foundation models,'' \emph{IEEE Trans. Syst. Man Cybern. Syst.}, vol.~53, no.~4, pp. 2037--2046, 2022.

\bibitem{wang2023new}
F.-Y. Wang, ``New control paradigm for industry 5.0: From big models to foundation control and management,'' \emph{IEEE/CAA J. of Autom. Sin.}, vol.~10, no.~8, pp. 1643--1646, 2023.

\bibitem{wang2023chat}
F.-Y. Wang, J.~Yang, X.~Wang, J.~Li, and Q.-L. Han, ``Chat with {ChatGPT} on industry 5.0: Learning and decision-making for intelligent industries,'' \emph{IEEE/CAA J. of Autom. Sin.}, vol.~10, no.~4, pp. 831--834, 2023.

\bibitem{qi2019smart}
Q.~Qi and F.~Tao, ``A smart manufacturing service system based on edge computing, fog computing, and cloud computing,'' \emph{IEEE Access}, vol.~7, pp. 86\,769--86\,777, 2019.

\bibitem{wang2021smart}
B.~Wang, F.~Tao, X.~Fang, C.~Liu, Y.~Liu, and T.~Freiheit, ``Smart manufacturing and intelligent manufacturing: A comparative review,'' \emph{Engineering}, vol.~7, no.~6, pp. 738--757, 2021.

\bibitem{garcia2022towards}
{\'A}.~Garc{\'\i}a, A.~Bregon, and M.~A. Mart{\'\i}nez-Prieto, ``Towards a connected digital twin learning ecosystem in manufacturing: Enablers and challenges,'' \emph{Comput. Industr. Eng.}, vol. 171, p. 108463, 2022.

\bibitem{andronie2021sustainable}
M.~Andronie, G.~L{\u{a}}z{\u{a}}roiu, R.~Ștef{\u{a}}nescu, C.~Uț{\u{a}}, and I.~Dijm{\u{a}}rescu, ``Sustainable, smart, and sensing technologies for cyber-physical manufacturing systems: A systematic literature review,'' \emph{Sustainability}, vol.~13, no.~10, p. 5495, 2021.

\bibitem{jiang2020review}
Y.~Jiang, S.~Yin, J.~Dong, and O.~Kaynak, ``A review on soft sensors for monitoring, control, and optimization of industrial processes,'' \emph{IEEE Sens. J.}, vol.~21, no.~11, pp. 12\,868--12\,881, 2020.

\bibitem{majid2022applications}
M.~Majid, S.~Habib, A.~R. Javed, M.~Rizwan, G.~Srivastava, T.~R. Gadekallu, and J.~C.-W. Lin, ``Applications of wireless sensor networks and internet of things frameworks in the industry revolution 4.0: A systematic literature review,'' \emph{Sensors}, vol.~22, no.~6, p. 2087, 2022.

\bibitem{aceto2019survey}
G.~Aceto, V.~Persico, and A.~Pescap{\'e}, ``A survey on information and communication technologies for industry 4.0: State-of-the-art, taxonomies, perspectives, and challenges,'' \emph{IEEE Commun. Surv. Tutor.}, vol.~21, no.~4, pp. 3467--3501, 2019.

\bibitem{colombo202170}
A.~W. Colombo, S.~Karnouskos, X.~Yu, O.~Kaynak, R.~C. Luo, Y.~Shi, P.~Leitao, L.~Ribeiro, and J.~Haase, ``A 70-year industrial electronics society evolution through industrial revolutions: The rise and flourishing of information and communication technologies,'' \emph{IEEE Ind. Electron. Mag.}, vol.~15, no.~1, pp. 115--126, 2021.

\bibitem{yang2023parallel}
J.~Yang, X.~Wang, Y.~Tian, and F.-Y. Wang, ``Parallel intelligence in {CPSSs}: Being, becoming, and believing,'' \emph{IEEE Intell. Syst.}, vol.~38, no.~06, pp. 75--80, 2023.

\bibitem{yang2023parallelCPS}
J.~Yang, X.~Wang, Y.~Wang \emph{et~al.}, ``Parallel intelligence and {CPSS} in 30 years: An {ACP} approach,'' \emph{Acta Automa. Sin.}, vol.~49, no.~3, pp. 614--634, 2023.

\bibitem{9896794}
X.~Wang, J.~Yang, J.~Han, W.~Wang, and F.-Y. Wang, ``Metaverses and {DeMetaverses}: From digital twins in {CPS} to parallel intelligence in {CPSS},'' \emph{IEEE Intell. Syst.}, vol.~37, no.~4, pp. 97--102, 2022.

\bibitem{wang2021survey}
Y.~Wang, ``Survey on deep multi-modal data analytics: Collaboration, rivalry, and fusion,'' \emph{ACM Trans. Multimedia Comput. Commun. Appl.}, vol.~17, no.~1s, pp. 1--25, 2021.

\bibitem{wu2023mfir}
L.~Wu, Y.~Long, C.~Gao, Z.~Wang, and Y.~Zhang, ``{MFIR}: Multimodal fusion and inconsistency reasoning for explainable fake news detection,'' \emph{Inf. Fusion}, vol. 100, p. 101944, 2023.

\bibitem{huang2023multi}
Z.~Huang, S.~Sun, J.~Zhao, and L.~Mao, ``Multi-modal policy fusion for end-to-end autonomous driving,'' \emph{Inf. Fusion}, vol.~98, p. 101834, 2023.

\bibitem{tian2023acf}
Y.~Tian, X.~Zhang, X.~Wang, J.~Xu, J.~Wang, R.~Ai, W.~Gu, and W.~Ding, ``{ACF-Net}: Asymmetric cascade fusion for {3D} detection with lidar point clouds and images,'' \emph{IEEE Trans. Intell. Veh.}, 2023.

\bibitem{wang2015social}
{F.-Y. Wang, X. Wang, Y. Yuan, T. Wang, Y. Lin}, ``Social computing and computational societies: The foundation and consequence of smart societies,'' \emph{Chinese Science Bulletin}, vol.~60, pp. 460--469, 2015.

\bibitem{adams2003semantic}
W.~Adams, G.~Iyengar, C.-Y. Lin, M.~R. Naphade, C.~Neti, H.~J. Nock, and J.~R. Smith, ``Semantic indexing of multimedia content using visual, audio, and text cues,'' \emph{EURASIP J. Adv. Signal Process.}, vol. 2003, pp. 1--16, 2003.

\bibitem{bourlard1996mew}
H.~Bourlard and S.~Dupont, ``A new {ASR} approach based on independent processing and recombination of partial frequency bands,'' in \emph{Int. Conf. Spoken Lang. Process. ICSLP Proc.}, vol.~1, 1996, pp. 426--429.

\bibitem{dogan2021machine}
A.~Dogan and D.~Birant, ``Machine learning and data mining in manufacturing,'' \emph{Expert Syst. Appl.}, vol. 166, p. 114060, 2021.

\bibitem{QU2023102172}
Z.~Qu, Y.~Meng, G.~Muhammad, and P.~Tiwari, ``{QMFND}: A quantum multimodal fusion-based fake news detection model for social media,'' \emph{Inf. Fusion}, p. 102172, 2023.

\bibitem{summaira2021recent}
J.~Summaira, X.~Li, A.~M. Shoib, S.~Li, and J.~Abdul, ``Recent advances and trends in multimodal deep learning: A review,'' \emph{arXiv preprint arXiv:2105.11087}, 2021.

\bibitem{antol2015vqa}
S.~Antol, A.~Agrawal, J.~Lu, M.~Mitchell, D.~Batra, C.~L. Zitnick, and D.~Parikh, ``{VQA}: Visual question answering,'' in \emph{Proc. IEEE Int. Conf. Comput. Vision}, 2015, pp. 2425--2433.

\bibitem{plummer2015flickr30k}
B.~A. Plummer, L.~Wang, C.~M. Cervantes, J.~C. Caicedo, J.~Hockenmaier, and S.~Lazebnik, ``Flickr30k entities: Collecting region-to-phrase correspondences for richer image-to-sentence models,'' in \emph{Proc. IEEE Int. Conf. Comput. Vision}, 2015, pp. 2641--2649.

\bibitem{cai2023multi}
Y.~Cai, X.~Sui, and G.~Gu, ``Multi-modal multi-task feature fusion for rgbt tracking,'' \emph{Inf. Fusion}, vol.~97, p. 101816, 2023.

\bibitem{vaswani2017attention}
A.~Vaswani, N.~Shazeer, N.~Parmar, J.~Uszkoreit, L.~Jones, A.~N. Gomez, {\L}.~Kaiser, and I.~Polosukhin, ``Attention is all you need,'' \emph{Adv. Neural Inf. Process. Syst.}, vol.~30, 2017.

\bibitem{devlin2018bert}
J.~Devlin, M.-W. Chang, K.~Lee, and K.~Toutanova, ``{BERT}: Pre-training of deep bidirectional {Transformers} for language understanding,'' in \emph{Conf. N. Am. Chapter Assoc. Comput. Linguistics: Hum. Lang. Technol.}, 2019, pp. 4171--4186.

\bibitem{brown2020language}
T.~Brown, B.~Mann, N.~Ryder, M.~Subbiah, J.~D. Kaplan, P.~Dhariwal, A.~Neelakantan, P.~Shyam, G.~Sastry, A.~Askell \emph{et~al.}, ``Language models are few-shot learners,'' \emph{Adv. Neural Inf. Process. Syst.}, vol.~33, pp. 1877--1901, 2020.

\bibitem{dosovitskiy2020image}
A.~Dosovitskiy, L.~Beyer, A.~Kolesnikov, D.~Weissenborn, X.~Zhai, T.~Unterthiner, M.~Dehghani, M.~Minderer, G.~Heigold, S.~Gelly \emph{et~al.}, ``An image is worth 16x16 words: Transformers for image recognition at scale,'' in \emph{Int. Conf. Learn. Represent.}, 2020.

\bibitem{carion2020end}
N.~Carion, F.~Massa, G.~Synnaeve, N.~Usunier, A.~Kirillov, and S.~Zagoruyko, ``End-to-end object detection with {Transformers},'' in \emph{Lect. Notes Comput. Sci.}\hskip 1em plus 0.5em minus 0.4em\relax Springer, 2020, pp. 213--229.

\bibitem{hu2021unit}
R.~Hu and A.~Singh, ``{UniT}: Multimodal multitask learning with a unified {Transformer},'' in \emph{Proc. IEEE Int. Conf. Comput. Vision}, 2021, pp. 1439--1449.

\bibitem{nguyen2023multimodal}
V.-A. Nguyen and S.~G. Kong, ``Multimodal feature fusion for illumination-invariant recognition of abnormal human behaviors,'' \emph{Inf. Fusion}, vol. 100, p. 101949, 2023.

\bibitem{raffel2020exploring}
C.~Raffel, N.~Shazeer, A.~Roberts, K.~Lee, S.~Narang, M.~Matena, Y.~Zhou, W.~Li, and P.~J. Liu, ``Exploring the limits of transfer learning with a unified text-to-text {Transformer},'' \emph{J. Mach. Learn. Res.}, vol.~21, no.~1, pp. 5485--5551, 2020.

\bibitem{lu202012}
J.~Lu, V.~Goswami, M.~Rohrbach, D.~Parikh, and S.~Lee, ``12-in-1: Multi-task vision and language representation learning,'' in \emph{Proc. IEEE Comput. Soc. Conf. Comput. Vision Pattern Recognit.}, 2020, pp. 10\,437--10\,446.

\bibitem{yoon2023multimedia}
J.~Yoon, G.~Choi, and C.~Choi, ``Multimedia analysis of robustly optimized multimodal transformer based on vision and language co-learning,'' \emph{Inf. Fusion}, vol. 100, p. 101922, 2023.

\bibitem{dou2022coarse}
Z.-Y. Dou, A.~Kamath, Z.~Gan, P.~Zhang, J.~Wang, L.~Li, Z.~Liu, C.~Liu, Y.~LeCun, N.~Peng \emph{et~al.}, ``Coarse-to-fine vision-language pre-training with fusion in the backbone,'' \emph{Adv. Neural Inf. Process. Syst.}, vol.~35, pp. 32\,942--32\,956, 2022.

\bibitem{singh2022flava}
A.~Singh, R.~Hu, V.~Goswami, G.~Couairon, W.~Galuba, M.~Rohrbach, and D.~Kiela, ``{FLAVA}: A foundational language and vision alignment model,'' in \emph{Proc. IEEE Comput. Soc. Conf. Comput. Vision Pattern Recognit.}, 2022, pp. 15\,638--15\,650.

\bibitem{wang2023one}
P.~Wang, S.~Wang, J.~Lin, S.~Bai, X.~Zhou, J.~Zhou, X.~Wang, and C.~Zhou, ``{ONE-PEACE}: Exploring one general representation model toward unlimited modalities,'' \emph{arXiv preprint arXiv:2305.11172}, 2023.

\bibitem{alayrac2022flamingo}
J.-B. Alayrac, J.~Donahue, P.~Luc, A.~Miech, I.~Barr, Y.~Hasson, K.~Lenc, A.~Mensch, K.~Millican, M.~Reynolds \emph{et~al.}, ``Flamingo: A visual language model for few-shot learning,'' \emph{Adv. Neural Inf. Process. Syst.}, vol.~35, pp. 23\,716--23\,736, 2022.

\bibitem{awadalla2023openflamingo}
A.~Awadalla, I.~Gao, J.~Gardner, J.~Hessel, Y.~Hanafy, W.~Zhu, K.~Marathe, Y.~Bitton, S.~Gadre, S.~Sagawa \emph{et~al.}, ``{OpenFlamingo}: An open-source framework for training large autoregressive vision-language models,'' \emph{arXiv preprint arXiv:2308.01390}, 2023.

\bibitem{gao2023llama}
P.~Gao, J.~Han, R.~Zhang, Z.~Lin, S.~Geng, A.~Zhou, W.~Zhang, P.~Lu, C.~He, X.~Yue \emph{et~al.}, ``{LLaMA-Adapter V2}: Parameter-efficient visual instruction model,'' \emph{arXiv preprint arXiv:2304.15010}, 2023.

\bibitem{bommasani2021opportunities}
R.~Bommasani, D.~A. Hudson, E.~Adeli, R.~Altman, S.~Arora, S.~von Arx, M.~S. Bernstein, J.~Bohg, A.~Bosselut, E.~Brunskill \emph{et~al.}, ``On the opportunities and risks of foundation models,'' \emph{arXiv preprint arXiv:2108.07258}, 2021.

\bibitem{li2023multimodal}
C.~Li, Z.~Gan, Z.~Yang, J.~Yang, L.~Li, L.~Wang, and J.~Gao, ``Multimodal foundation models: From specialists to general-purpose assistants,'' \emph{arXiv preprint arXiv:2309.10020}, vol.~1, no.~2, 2023.

\bibitem{xu2023multimodal}
P.~Xu, X.~Zhu, and D.~A. Clifton, ``Multimodal learning with transformers: A survey,'' \emph{IEEE Trans. Pattern Anal. Mach. Intell.}, 2023.

\bibitem{wang2023large}
X.~Wang, G.~Chen, G.~Qian, P.~Gao, X.-Y. Wei, Y.~Wang, Y.~Tian, and W.~Gao, ``Large-scale multi-modal pre-trained models: A comprehensive survey,'' \emph{Mach. Intell. Res.}, pp. 1--36, 2023.

\bibitem{kuang2014tensor}
L.~Kuang, F.~Hao, L.~T. Yang, M.~Lin, C.~Luo, and G.~Min, ``A tensor-based approach for big data representation and dimensionality reduction,'' \emph{IEEE Trans. Emerg. Topics Comput.}, vol.~2, no.~3, pp. 280--291, 2014.

\bibitem{kuang2016tensor}
L.~Kuang, L.~T. Yang, and K.~Qiu, ``Tensor-based software-defined internet of things,'' \emph{IEEE Wirel. Commun.}, vol.~23, no.~5, pp. 84--89, 2016.

\bibitem{7134729}
L.~Kuang, L.~T. Yang, J.~Chen, F.~Hao, and C.~Luo, ``A holistic approach for distributed dimensionality reduction of big data,'' \emph{IEEE Trans. on Cloud Comput.}, vol.~6, no.~2, pp. 506--518, 2018.

\bibitem{7364232}
L.~Kuang, L.~T. Yang, and Y.~Liao, ``An integration framework on cloud for cyber-physical-social systems big data,'' \emph{IEEE Trans. on Cloud Comput.}, vol.~8, no.~2, pp. 363--374, 2020.

\bibitem{abed2022contemporary}
K.~Abed-Meraim, N.~L. Trung, A.~Hafiane \emph{et~al.}, ``A contemporary and comprehensive survey on streaming tensor decomposition,'' \emph{IEEE Trans. Knowl. Data Eng.}, 2022.

\bibitem{kolda2009tensor}
T.~G. Kolda and B.~W. Bader, ``Tensor decompositions and applications,'' \emph{SIAM Rev.}, vol.~51, no.~3, pp. 455--500, 2009.

\bibitem{wang2019ho}
P.~Wang, L.~T. Yang, G.~Qian, J.~Li, and Z.~Yan, ``{HO-OTSVD}: A novel tensor decomposition and its incremental decomposition for cyber-physical-social networks ({CPSN}),'' \emph{IEEE Trans. Netw. Sci. Eng.}, vol.~7, no.~2, pp. 713--725, 2019.

\bibitem{feng2018privacy}
J.~Feng, L.~T. Yang, Q.~Zhu, and K.-K.~R. Choo, ``Privacy-preserving tensor decomposition over encrypted data in a federated cloud environment,'' \emph{IEEE Trans. Dependable Secure Comput.}, vol.~17, no.~4, pp. 857--868, 2018.

\bibitem{feng2019practical}
J.~Feng, L.~T. Yang, and R.~Zhang, ``Practical privacy-preserving high-order bi-lanczos in integrated edge-fog-cloud architecture for cyber-physical-social systems,'' \emph{ACM Trans. Internet Technol.}, vol.~19, no.~2, pp. 1--18, 2019.

\bibitem{wang2018cyber}
P.~Wang, L.~T. Yang, G.~Qian, and F.~Lu, ``The cyber-physical-social transition tensor service framework,'' \emph{IEEE Trans. Sustain. Comput.}, vol.~6, no.~3, pp. 481--492, 2018.

\bibitem{8701474}
P.~Wang, L.~T. Yang, Y.~Peng, J.~Li, and X.~Xie, ``${M^2}{T^2}$: The multivariate multistep transition tensor for user mobility pattern prediction,'' \emph{IEEE Trans. Netw. Sci. Eng.}, vol.~7, no.~2, pp. 907--917, 2020.

\bibitem{wang2018edge}
P.~Wang, L.~T. Yang, and J.~Li, ``An edge cloud-assisted {CPSS} framework for smart city,'' \emph{IEEE Cloud Comput.}, vol.~5, no.~5, pp. 37--46, 2018.

\bibitem{9208706}
H.~Liu, L.~T. Yang, T.~Yao, J.~Ding, and A.~Deng, ``Tensor-train-based higher order dominant {Z}-eigen decomposition for multi-modal prediction and its cloud/edge implementation,'' \emph{IEEE Trans. Netw. Sci. Eng.}, vol.~8, no.~2, pp. 1353--1366, 2021.

\bibitem{8449102}
H.~Liu, L.~T. Yang, Y.~Guo, X.~Xie, and J.~Ma, ``An incremental tensor-train decomposition for cyber-physical-social big data,'' \emph{IEEE Trans. Big Data}, vol.~7, no.~2, pp. 341--354, 2021.

\bibitem{liu2022tensor}
H.~Liu, J.~Wang, X.~Yin, J.~Ding, L.~T. Yang, T.~Yao, J.~Yang, and Y.~Gao, ``Tensor-train-based multiuser multivariate multiorder physical markov process informed multimodal prediction for industrial trajectory applications,'' \emph{IEEE Trans. Industr. Inform.}, 2022.

\bibitem{Yu2020extra}
L.~Yu, Z.~Guo, G.~Chen, and Y.~Xi, ``Summary of knowledge graph construction oriented knowledge extraction technology,'' \emph{J. Inf. Eng. Univ.}, vol.~21, no.~2, pp. 227--235, 2020.

\bibitem{zhuang2023knowledge}
L.~Zhuang, H.~Fei, and P.~Hu, ``Knowledge-enhanced event relation extraction via event ontology prompt,'' \emph{Inf. Fusion}, vol. 100, p. 101919, 2023.

\bibitem{segura2011linguistic}
I.~Segura-Bedmar, P.~Mart{\'\i}nez, and C.~de~Pablo-S{\'a}nchez, ``A linguistic rule-based approach to extract drug-drug interactions from pharmacological documents,'' in \emph{BMC Bioinform.}, vol.~12, no.~2, 2011, pp. 1--11.

\bibitem{pawar2020extracting}
S.~Pawar, P.~Bhattacharyya, and G.~K. Palshikar, ``Extracting {N}-ary cross-sentence relations using constrained subsequence kernel,'' \emph{arXiv preprint arXiv:2006.08185}, 2020.

\bibitem{xu2023generic}
Y.~Xu, L.~Han, T.~Zhu, L.~Sun, B.~Du, and W.~Lv, ``Generic dynamic graph convolutional network for traffic flow forecasting,'' \emph{Inf. Fusion}, vol. 100, p. 101946, 2023.

\bibitem{fu2019graphrel}
T.-J. Fu, P.-H. Li, and W.-Y. Ma, ``{GraphRel}: Modeling text as relational graphs for joint entity and relation extraction,'' in \emph{Annu. Meet. Assoc. Comput. Linguist., Proc. Conf.}, 2019, pp. 1409--1418.

\bibitem{Li2020ASO}
C.~Li, A.~Li, Y.~Wang, H.~Tu, and Y.~Song, ``A survey on approaches and applications of knowledge representation learning,'' \emph{IEEE Int. Conf. Data Sci. Cyberspace, DSC}, pp. 312--319, 2020.

\bibitem{bordes2013translating}
A.~Bordes, N.~Usunier, A.~Garcia-Duran, J.~Weston, and O.~Yakhnenko, ``Translating embeddings for modeling multi-relational data,'' \emph{Adv. Neural Inf. Process. Syst.}, vol.~26, 2013.

\bibitem{wang2014knowledge}
Z.~Wang, J.~Zhang, J.~Feng, and Z.~Chen, ``Knowledge graph embedding by translating on hyperplanes,'' in \emph{Proc. Natl. Conf. Artif. Intell.}, vol.~28, no.~1, 2014.

\bibitem{ji2015knowledge}
G.~Ji, S.~He, L.~Xu, K.~Liu, and J.~Zhao, ``Knowledge graph embedding via dynamic mapping matrix,'' in \emph{Annu. Meet. Assoc. Comput. Linguist. Int. Jt. Conf. Nat. Lang. Process. Asian Fed. Nat. Lang. Process., Proc. Conf.}, 2015, pp. 687--696.

\bibitem{nickel2011three}
M.~Nickel, V.~Tresp, H.-P. Kriegel \emph{et~al.}, ``A three-way model for collective learning on multi-relational data,'' in \emph{Proc. Int. Conf. Mach. Learn., ICML}, vol.~11, no. 10.5555, 2011, pp. 3\,104\,482--3\,104\,584.

\bibitem{yang2014embedding}
B.~Yang, W.-t. Yih, X.~He, J.~Gao, and L.~Deng, ``Embedding entities and relations for learning and inference in knowledge bases,'' in \emph{Int. Conf. Learn. Represent., ICLR}, 2015.

\bibitem{kazemi2018simple}
S.~M. Kazemi and D.~Poole, ``Simple embedding for link prediction in knowledge graphs,'' \emph{Adv. Neural Inf. Process. Syst.}, vol.~31, 2018.

\bibitem{trouillon2016complex}
T.~Trouillon, J.~Welbl, S.~Riedel, {\'E}.~Gaussier, and G.~Bouchard, ``Complex embeddings for simple link prediction,'' in \emph{Int. Conf. Mach. Learn., ICML}, 2016, pp. 2071--2080.

\bibitem{nickel2016holographic}
M.~Nickel, L.~Rosasco, and T.~Poggio, ``Holographic embeddings of knowledge graphs,'' in \emph{AAAI Conf. Artif. Intell., AAAI}, vol.~30, no.~1, 2016.

\bibitem{bordes2014semantic}
A.~Bordes, X.~Glorot, J.~Weston, and Y.~Bengio, ``A semantic matching energy function for learning with multi-relational data: Application to word-sense disambiguation,'' \emph{Mach. Learn.}, vol.~94, pp. 233--259, 2014.

\bibitem{dettmers2018convolutional}
T.~Dettmers, P.~Minervini, P.~Stenetorp, and S.~Riedel, ``Convolutional {2D} knowledge graph embeddings,'' in \emph{AAAI Conf. Artif. Intell., AAAI}, vol.~32, no.~1, 2018.

\bibitem{nguyen2017novel}
D.~Q. Nguyen, T.~D. Nguyen, D.~Q. Nguyen, and D.~Phung, ``A novel embedding model for knowledge base completion based on convolutional neural network,'' in \emph{Conf. N. Am. Chapter Assoc. Comput. Linguistics: Hum. Lang. Technol.}, vol.~2, 2018.

\bibitem{schlichtkrull2018modeling}
M.~Schlichtkrull, T.~N. Kipf, P.~Bloem, R.~Van Den~Berg, I.~Titov, and M.~Welling, ``Modeling relational data with graph convolutional networks,'' in \emph{Lect. Notes Comput. Sci.}, vol. 10843, 2018, pp. 593--607.

\bibitem{wang2019robust}
Z.~Wang, Z.~Ren, C.~He, P.~Zhang, and Y.~Hu, ``Robust embedding with multi-level structures for link prediction,'' in \emph{IJCAI Int. Joint Conf. Artif. Intell.}, 2019, pp. 5240--5246.

\bibitem{nathani2019learning}
D.~Nathani, J.~Chauhan, C.~Sharma, and M.~Kaul, ``Learning attention-based embeddings for relation prediction in knowledge graphs,'' in \emph{Annu. Meet. Assoc. Comput. Linguist., Proc. Conf.}, 2020.

\bibitem{wang2019logic}
P.~Wang, J.~Han, C.~Li, and R.~Pan, ``Logic attention based neighborhood aggregation for inductive knowledge graph embedding,'' in \emph{AAAI Conf. Artif. Intell., AAAI}, vol.~33, no.~01, 2019, pp. 7152--7159.

\bibitem{Chang2017knowledge}
L.~Chang, M.~Zhu, T.~Gu, C.~Bin, J.~Qian, and J.~Zhang, ``Knowledge graph embedding by dynamic translation,'' \emph{IEEE Access}, vol.~5, pp. 20\,898--20\,907, 2017.

\bibitem{yue2023knowlenet}
T.~Yue, R.~Mao, H.~Wang, Z.~Hu, and E.~Cambria, ``{KnowleNet}: Knowledge fusion network for multimodal sarcasm detection,'' \emph{Inf. Fusion}, vol. 100, p. 101921, 2023.

\bibitem{wang2022multi}
Z.~Wang, X.~He, H.~Shen, S.~Fan, and Y.~Zeng, ``Multi-source information fusion to identify water supply pipe leakage based on {SVM} and {VMD},'' \emph{Inf. Process. Manag.}, vol.~59, no.~2, p. 102819, 2022.

\bibitem{lehmann2015dbpedia}
J.~Lehmann, R.~Isele, M.~Jakob, A.~Jentzsch, D.~Kontokostas, P.~N. Mendes, S.~Hellmann, M.~Morsey, P.~Van~Kleef, S.~Auer \emph{et~al.}, ``{DBpedia}--a large-scale, multilingual knowledge base extracted from {Wikipedia},'' \emph{Semant. Web}, vol.~6, no.~2, pp. 167--195, 2015.

\bibitem{suchanek2011paris}
F.~M. Suchanek, S.~Abiteboul, and P.~Senellart, ``{PARIS}: Probabilistic alignment of relations, instances, and schema,'' \emph{Proc. VLDB Endow.}, vol.~5, no.~3, pp. 157--168, 2011.

\bibitem{shao2016rimom}
C.~Shao, L.-M. Hu, J.-Z. Li, Z.-C. Wang, T.~Chung, and J.-B. Xia, ``{RiMOM-IM}: A novel iterative framework for instance matching,'' \emph{J. Comput. Sci. Technol.}, vol.~31, pp. 185--197, 2016.

\bibitem{Zhang2020entity}
F.~Zhang, L.~Yang, J.~Li, and J.-W. Cheng, ``An overview of entity alignment methods,'' \emph{Chinese J. Comput.}, vol.~45, no.~6, pp. 1195--1225, 2020.

\bibitem{Chen2016MultilingualKG}
M.~Chen, Y.~Tian, M.~Yang, and C.~Zaniolo, ``Multilingual knowledge graph embeddings for cross-lingual knowledge alignment,'' in \emph{IJCAI Int. Joint Conf. Artif. Intell.}, 2016.

\bibitem{Zhu2017IterativeEA}
H.~Zhu, R.~Xie, Z.~Liu, and M.~Sun, ``Iterative entity alignment via joint knowledge embeddings,'' in \emph{IJCAI Int. Joint Conf. Artif. Intell.}, vol.~17, 2017, pp. 4258--4264.

\bibitem{Sun2018BootstrappingEA}
Z.~Sun, W.~Hu, Q.~Zhang, and Y.~Qu, ``Bootstrapping entity alignment with knowledge graph embedding,'' in \emph{IJCAI Int. Joint Conf. Artif. Intell.}, vol.~18, no. 2018, 2018.

\bibitem{Zhu2019NeighborhoodAwareAR}
Q.~Zhu, X.~Zhou, J.~Wu, J.~Tan, and L.~Guo, ``Neighborhood-aware attentional representation for multilingual knowledge graphs,'' in \emph{IJCAI Int. Joint Conf. Artif. Intell.}, 2019, pp. 1943--1949.

\bibitem{Sun2019TransEdgeTR}
Z.~Sun, J.~Huang, W.~Hu, M.~Chen, L.~Guo, and Y.~Qu, ``{TransEdge}: Translating relation-contextualized embeddings for knowledge graphs,'' in \emph{Lect. Notes Comput. Sci.}, vol. 11778, 2019, pp. 612--629.

\bibitem{Lin2019GuidingCE}
X.~Lin, H.~Yang, J.~Wu, C.~Zhou, and B.~Wang, ``Guiding cross-lingual entity alignment via adversarial knowledge embedding,'' in \emph{Proc. IEEE Int. Conf. Data Min. ICDM}, 2019, pp. 429--438.

\bibitem{Chen2018CotrainingEO}
M.~Chen, Y.~Tian, K.-W. Chang, S.~Skiena, and C.~Zaniolo, ``Co-training embeddings of knowledge graphs and entity descriptions for cross-lingual entity alignment,'' in \emph{IJCAI Int. Joint Conf. Artif. Intell.}, 2018, pp. 3998--4004.

\bibitem{Wang2018CrosslingualKG}
Z.~Wang, Q.~Lv, X.~Lan, and Y.~Zhang, ``Cross-lingual knowledge graph alignment via graph convolutional networks,'' in \emph{Proc. Conf. Empir. Methods Nat. Lang. Process., EMNLP}, 2018, pp. 349--357.

\bibitem{Wu2019JointlyLE}
Y.~Wu, X.~Liu, Y.~Feng, Z.~Wang, and D.~Zhao, ``Jointly learning entity and relation representations for entity alignment,'' in \emph{Conf. Empir. Methods Nat. Lang. Process. Int. Jt. Conf. Nat. Lang. Process.}, 2019, pp. 240--249.

\bibitem{Li2019SemisupervisedEA}
C.~Li, Y.~Cao, L.~Hou, J.~Shi, J.-Z. Li, and T.-S. Chua, ``Semi-supervised entity alignment via joint knowledge embedding model and cross-graph model,'' in \emph{Conf. Empir. Methods Nat. Lang. Process. Int. Jt. Conf. Nat. Lang. Process.}, 2019, pp. 2723--2732.

\bibitem{Sun2019KnowledgeGA}
Z.~Sun, C.~Wang, W.~Hu, M.~Chen, J.~Dai, W.~Zhang, and Y.~Qu, ``Knowledge graph alignment network with gated multi-hop neighborhood aggregation,'' in \emph{AAAI Conf. Artif. Intell.}, vol.~34, no.~01, 2020, pp. 222--229.

\bibitem{Xie2016ImageembodiedKR}
R.~Xie, Z.~Liu, H.~Luan, and M.~Sun, ``Image-embodied knowledge representation learning,'' in \emph{IJCAI Int. Joint Conf. Artif. Intell.}, 2016, pp. 3140--3146.

\bibitem{dong2023knowledge}
W.~Dong, S.~Sun, J.~Zhao, and N.~Zhang, ``Knowledge graph relation reasoning with variational reinforcement network,'' \emph{Inf. Fusion}, vol. 100, p. 101900, 2023.

\bibitem{Gardner2014IncorporatingVS}
M.~Gardner, P.~P. Talukdar, J.~Krishnamurthy, and T.~M. Mitchell, ``Incorporating vector space similarity in random walk inference over knowledge bases,'' in \emph{Conf. Empir. Methods Nat. Lang. Process.}, 2014, pp. 397--406.

\bibitem{Neelakantan2015CompositionalVS}
A.~Neelakantan, B.~Roth, and A.~McCallum, ``Compositional vector space models for knowledge base completion,'' in \emph{Annu. Meet. Assoc. Comput. Linguist. Int. Jt. Conf. Nat. Lang. Process. Asian Fed. Nat. Lang. Process.}, 2015, pp. 156--166.

\bibitem{Xiong2017DeepPathAR}
W.~Xiong, T.-L.-G. Hoang, and W.~Y. Wang, ``{DeepPath}: A reinforcement learning method for knowledge graph reasoning,'' in \emph{Conf. Empir. Methods Nat. Lang. Process.}, 2017, pp. 564--573.

\bibitem{Galrraga2013AMIEAR}
L.~A. Gal{\'a}rraga, C.~Teflioudi, K.~Hose, and F.~Suchanek, ``{AMIE}: Association rule mining under incomplete evidence in ontological knowledge bases,'' in \emph{Proc. Int. Conf. World Wide Web}, 2013, pp. 413--422.

\bibitem{Guo2016JointlyEK}
S.~Guo, Q.~Wang, L.~Wang, B.~Wang, and L.~Guo, ``Jointly embedding knowledge graphs and logical rules,'' in \emph{Conf. Empir. Methods Nat. Lang. Process.}, 2016, pp. 192--202.

\bibitem{Guo2017KnowledgeGE}
{S. Guo, Q. Wang, L. Wang, B. Wang, and L. Guo}, ``Knowledge graph embedding with iterative guidance from soft rules,'' in \emph{AAAI Conf. Artif. Intell.}, vol.~32, no.~1, 2018.

\bibitem{Trivedi2017KnowEvolveDT}
R.~S. Trivedi, H.~Dai, Y.~Wang, and L.~Song, ``{Know-Evolve}: Deep temporal reasoning for dynamic knowledge graphs,'' in \emph{Int. Conf. Mach. Learn., ICML}, vol.~7, 2017, pp. 3462--3471.

\bibitem{GarcaDurn2018LearningSE}
A.~Garc{\'i}a-Dur{\'a}n, S.~Dumancic, and M.~Niepert, ``Learning sequence encoders for temporal knowledge graph completion,'' in \emph{Proc. Conf. Empir. Methods Nat. Lang. Process.}, 2018, pp. 4816--4821.

\bibitem{Wu2020TeMPTM}
J.~Wu, M.~Cao, J.~C.~K. Cheung, and W.~L. Hamilton, ``{TeMP}: Temporal message passing for temporal knowledge graph completion,'' in \emph{Conf. Empir. Methods Nat. Lang. Process.}, 2020, pp. 5730--5746.

\bibitem{Zhu2020LearningFH}
C.~Zhu, M.~Chen, C.~Fan, G.~Cheng, and Y.~Zhan, ``Learning from history: Modeling temporal knowledge graphs with sequential copy-generation networks,'' in \emph{AAAI Conf. Artif. Intell.}, vol.~35, no.~5, 2021, pp. 4732--4740.

\bibitem{Liu2022DANetDA}
K.~Liu, F.~Zhao, H.~Chen, Y.~Li, G.~Xu, and H.~Jin, ``{DA-Net}: Distributed attention network for temporal knowledge graph reasoning,'' in \emph{Int. Conf. Inf. Knowledge Manage.}, 2022, pp. 1289--1298.

\bibitem{Zhang2016CollaborativeKB}
F.~Zhang, N.~J. Yuan, D.~Lian, X.~Xie, and W.-Y. Ma, ``Collaborative knowledge base embedding for recommender systems,'' in \emph{Proc. ACM SIGKDD Int. Conf. Knowl. Discov. Data Min.}, 2016, pp. 353--362.

\bibitem{Xie2016RepresentationLO}
R.~Xie, Z.~Liu, J.~Jia, H.~Luan, and M.~Sun, ``Representation learning of knowledge graphs with entity descriptions,'' in \emph{AAAI Conf. Artif. Intell.}, vol.~30, no.~1, 2016, pp. 2659--2665.

\bibitem{pan2023unifying}
S.~Pan, L.~Luo, Y.~Wang, C.~Chen, J.~Wang, and X.~Wu, ``Unifying large language models and knowledge graphs: A roadmap,'' \emph{arXiv preprint arXiv:2306.08302}, 2023.

\bibitem{wang2021kepler}
X.~Wang, T.~Gao, Z.~Zhu, Z.~Zhang, Z.~Liu, J.~Li, and J.~Tang, ``{KEPLER}: A unified model for knowledge embedding and pre-trained language representation,'' \emph{Trans. Assoc. Comput. Linguist.}, vol.~9, pp. 176--194, 2021.

\bibitem{ke2021jointgt}
P.~Ke, H.~Ji, Y.~Ran, X.~Cui, L.~Wang, L.~Song, X.~Zhu, and M.~Huang, ``{JointGT}: Graph-text joint representation learning for text generation from knowledge graphs,'' in \emph{Find. Assoc. Comput. Linguist.: ACL-IJCNLP}, 2021, pp. 2526--2538.

\bibitem{yasunaga2022deep}
M.~Yasunaga, A.~Bosselut, H.~Ren, X.~Zhang, C.~D. Manning, P.~S. Liang, and J.~Leskovec, ``Deep bidirectional language-knowledge graph pretraining,'' \emph{Adv. Neural Inf. Process. Syst.}, vol.~35, pp. 37\,309--37\,323, 2022.

\bibitem{wang2004parallel}
F.-Y. Wang, ``Parallel system methods for management and control of complex systems,'' \emph{Control Decis.}, vol.~19, pp. 485--489, 2004.

\bibitem{wang2016steps}
F.-Y. Wang, X.~Wang, L.~Li, and L.~Li, ``Steps toward parallel intelligence,'' \emph{IEEE/CAA J. of Autom. Sin.}, vol.~3, no.~4, pp. 345--348, 2016.

\bibitem{cao2023parameter}
Y.~Cao, Y.~Wang, J.~Wang, Y.~Tian, X.~Wang, and F.-Y. Wang, ``Parameter identification and refinement for parallel {PCB} inspection in cyber--physical--social systems,'' \emph{IEEE Trans. Comput. Soc. Syst.}, 2023.

\bibitem{wangfeiyue2018Parallel}
F.-Y. Wang, Y.~Gao, X.~Shang \emph{et~al.}, ``Parallel manufacturing and industries 5.0: From virtual manufacturing to intelligent manufacturing,'' \emph{Sci.Technol. Rev.}, vol.~36, no.~21, pp. 10--22, 2018.

\bibitem{li2018Parallel}
L.~Li, X.~Wang, and X.~Shang, ``Parallel manufacturing for textile, footwear and garment industries,'' \emph{Sci. Technol. Rev.}, vol.~36, no.~21, pp. 48--55, 2018.

\bibitem{li2021distributed}
L.~Li, P.~Sun, and J.~Lu, ``Distributed manufacturing in knitting industry,'' in \emph{Proc. IEEE Int. Conf. Digit. Twins Parallel Intell.}\hskip 1em plus 0.5em minus 0.4em\relax IEEE, 2021, pp. 74--77.

\bibitem{yang2022parallel}
J.~Yang, X.~Wang, and Y.~Zhao, ``Parallel manufacturing for industrial metaverses: A new paradigm in smart manufacturing,'' \emph{IEEE/CAA J. of Autom. Sin.}, vol.~9, no.~12, pp. 2063--2070, 2022.

\bibitem{yang2022defact}
J.~Yang, S.~Li, X.~Wang, J.~Lu, H.~Wu, and X.~Wang, ``Defact in manuverse for parallel manufacturing: Foundation models and parallel workers in smart factories,'' \emph{IEEE Trans. Syst. Man Cybern. Syst.}, vol.~53, no.~4, pp. 2188--2199, 2022.

\bibitem{jogin2018feature}
M.~Jogin, M.~Madhulika, G.~Divya, R.~Meghana, S.~Apoorva \emph{et~al.}, ``Feature extraction using convolution neural networks (cnn) and deep learning,'' in \emph{IEEE Int. Conf. Recent Trends Electron., Inf. Commun. Technol.,}.\hskip 1em plus 0.5em minus 0.4em\relax IEEE, 2018, pp. 2319--2323.

\bibitem{tabassum2020survey}
A.~Tabassum and R.~R. Patil, ``A survey on text pre-processing \& feature extraction techniques in natural language processing,'' \emph{Int. Res. J. Eng. Technol.}, vol.~7, no.~06, pp. 4864--4867, 2020.

\bibitem{jaegle2021perceiver}
A.~Jaegle, F.~Gimeno, A.~Brock, O.~Vinyals, A.~Zisserman, and J.~Carreira, ``Perceiver: General perception with iterative attention,'' in \emph{Proc. Mach. Learn. Res.}, 2021, pp. 4651--4664.

\bibitem{yinglei2022smart}
H.~Yinglei, Q.~Dexin, and Z.~Shengyuan, ``Smart transportation travel model based on multiple data sources fusion for defense systems,'' \emph{Soft Comput.}, vol.~26, no.~7, pp. 3247--3259, 2022.

\bibitem{liu2022intelligent}
Q.~Liu, M.~Liu, H.~Zhou, F.~Yan, Y.~Ma, and W.~Shen, ``Intelligent manufacturing system with human-cyber-physical fusion and collaboration for process fine control,'' \emph{J. Manuf. Syst.}, vol.~64, pp. 149--169, 2022.

\bibitem{li2020acquisition}
Y.~Li, H.~Jia, J.~Qi, H.~Sun, X.~Tian, H.~Liu, and X.~Fan, ``An acquisition method of agricultural equipment roll angle based on multi-source information fusion,'' \emph{Sensors}, vol.~20, no.~7, p. 2082, 2020.

\bibitem{choudhary2020comprehensive}
T.~Choudhary, V.~Mishra, A.~Goswami, and J.~Sarangapani, ``A comprehensive survey on model compression and acceleration,'' \emph{Artif. Intell. Rev.}, vol.~53, pp. 5113--5155, 2020.

\bibitem{feng2020privacy}
J.~Feng, L.~T. Yang, N.~J. Gati, X.~Xie, and B.~S. Gavuna, ``Privacy-preserving computation in cyber-physical-social systems: A survey of the state-of-the-art and perspectives,'' \emph{Inf. Sci.}, vol. 527, pp. 341--355, 2020.

\bibitem{al2023privacy}
M.~Al-Hawawreh and M.~S. Hossain, ``A privacy-aware framework for detecting cyber attacks on internet of medical things systems using data fusion and quantum deep learning,'' \emph{Inf. Fusion}, p. 101889, 2023.

\bibitem{feng2023tensor}
J.~Feng, L.~T. Yang, B.~Ren, D.~Zou, M.~Dong, and S.~Zhang, ``Tensor recurrent neural network with differential privacy,'' \emph{IEEE Trans. Comput.}, 2023.

\bibitem{zhang2022task}
P.~Zhang, X.~Cheng, S.~Su, and N.~Wang, ``Task allocation under geo-indistinguishability via group-based noise addition,'' \emph{IEEE Trans. Big Data}, vol.~9, no.~3, pp. 860--877, 2022.

\bibitem{zhang2023effective}
------, ``Effective truth discovery under local differential privacy by leveraging noise-aware probabilistic estimation and fusion,'' \emph{Knowl. Based Syst.}, vol. 261, p. 110213, 2023.

\bibitem{ZHOU2023102182}
X.~Zhou, Q.~Yang, Q.~Liu, W.~Liang, K.~Wang, Z.~Liu, J.~Ma, and Q.~Jin, ``Spatial-temporal federated transfer learning with multi-sensor data fusion for cooperative positioning,'' \emph{Inf. Fusion}, p. 102182, 2023.

\bibitem{diaz2023connecting}
N.~D{\'\i}az-Rodr{\'\i}guez, J.~Del~Ser, M.~Coeckelbergh, M.~L. de~Prado, E.~Herrera-Viedma, and F.~Herrera, ``Connecting the dots in trustworthy artificial intelligence: From {AI} principles, ethics, and key requirements to responsible {AI} systems and regulation,'' \emph{Inf. Fusion}, p. 101896, 2023.

\bibitem{qi2023fl}
P.~Qi, D.~Chiaro, and F.~Piccialli, ``{FL-FD}: Federated learning-based fall detection with multimodal data fusion,'' \emph{Inf. Fusion}, p. 101890, 2023.

\bibitem{yang2023tensor}
L.~T. Yang, R.~Zhao, D.~Liu, W.~Lu, and X.~Deng, ``Tensor-empowered federated learning for cyber-physical-social computing and communication systems,'' \emph{IEEE Commun. Surv. Tutor.}, vol.~25, no.~3, pp. 1909--1940, 2023.

\bibitem{yuan2018blockchain}
Y.~Yuan and F.-Y. Wang, ``Blockchain and cryptocurrencies: Model, techniques, and applications,'' \emph{IEEE Trans. Syst. Man Cybern. Syst.}, vol.~48, no.~9, pp. 1421--1428, 2018.

\bibitem{Xu2023blockchain}
C.~Xu, H.~Wu, H.~Liu, W.~Gu, Y.~Li, and D.~Cao, ``Blockchain-oriented privacy protection of sensitive data in the internet of vehicles,'' \emph{IEEE Trans. Intell. Veh.}, vol.~8, no.~2, pp. 1057--1067, 2023.

\bibitem{Lang2022blockchain}
J.~Li, J.~Li \emph{et~al.}, ``Multi-blockchain based data trading markets with novel pricing mechanisms,'' \emph{IEEE/CAA J. of Autom. Sin.}, vol.~10, no.~12, pp. 2222–--2232, 2023.

\bibitem{wang2019blockchain}
S.~Wang, L.~Ouyang \emph{et~al.}, ``Blockchain-enabled smart contracts: architecture, applications, and future trends,'' \emph{IEEE Trans. Syst. Man Cybern. Syst.}, vol.~49, no.~11, pp. 2266--2277, 2019.

\bibitem{Leng2023Blockchain}
J.~Leng, X.~Zhu \emph{et~al.}, ``Manuchain ii: Blockchained smart contract system as the digital twin of decentralized autonomous manufacturing toward resilience in industry 5.0,'' \emph{IEEE Trans. Syst. Man Cybern. Syst.}, vol.~53, no.~8, pp. 4715--4728, 2023.

\bibitem{wang2018overview}
S.~Wang, Y.~Yuan \emph{et~al.}, ``An overview of smart contract: Architecture, applications, and future trends,'' in \emph{IEEE Intell. Veh. Symp. Proc.}\hskip 1em plus 0.5em minus 0.4em\relax IEEE, 2018, pp. 108--113.

\end{thebibliography}

\end{document}